\documentclass[12pt,preprint,useAMS,usenatbib]{emulateapj}
\usepackage{graphicx, txfonts}
\usepackage{setspace}
\usepackage{natbib}
\usepackage{url}
\usepackage{multirow}

\usepackage[usenames,dvipsnames]{color}

\newcommand{\AlII}{Al\,\textsc{ii}}
\newcommand{\AlIII}{Al\,\textsc{iii}}

\newcommand{\CI}{C\,\textsc{i}}
\newcommand{\CII}{C\,\textsc{ii}}

\newcommand{\CrII}{\textrm{Cr}\,\textsc{ii}}
\newcommand{\DI}{\textrm{D}\,\textsc{i}}
\newcommand{\FeII}{Fe\,\textsc{ii}}
\newcommand{\FeIII}{Fe\,\textsc{iii}}

\newcommand{\HI}{\textrm{H}\,\textsc{i}}

\newcommand{\HII}{\textrm{H}\,\textsc{ii}}

\newcommand{\Lya}{Ly$\alpha$}

\newcommand{\NI}{N\,\textsc{i}}
\newcommand{\NII}{N\,\textsc{ii}}

\newcommand{\NiII}{Ni\,\textsc{ii}}
\newcommand{\OI}{O\,\textsc{i}}

\newcommand{\SII}{S\,\textsc{ii}}
\newcommand{\SiII}{Si\,\textsc{ii}}
\newcommand{\SiIII}{Si\,\textsc{iii}}
\newcommand{\SiIV}{Si\,\textsc{iv}}

\newcommand{\ZnII}{\textrm{Zn}\,\textsc{ii}}

\newcommand*\xbar[1]{%
  \hbox{%
    \vbox{%
      \hrule height 0.5pt 
      \kern0.5ex
      \hbox{%
        \kern-0.1em
        \ensuremath{#1}%
        \kern-0.1em
      }%
    }%
  }%
}

\shorttitle{\textsc{Dwarf galaxies at high redshift}}
\shortauthors{\textsc{Cooke, Pettini, \& Jorgenson}}

\begin{document}

\title{The most metal-poor damped Lyman alpha systems:\\An insight into dwarf galaxies at high redshift\altaffilmark{$\star$}}

\author{Ryan J. Cooke\altaffilmark{1,4}, Max Pettini\altaffilmark{2} and Regina A. Jorgenson\altaffilmark{3}}

\altaffiltext{$\star$}{Based on observations collected at the European Organisation for Astronomical Research 
in the Southern Hemisphere, Chile [VLT program IDs:
60.A-9022(A), 65.O-0063(B), 65.O-0296(A), 67.A-0022(A), 67.A-0078(A), 68.A-0600(A), 68.B-0115(A), 70.A-0425(C), 078.A-0185(A), 080.A-0014(A), 082.A-0544(A), 083.A-0042(A), 083.A-0454(A), 085.A-0109(A), 086.A-0204(A)],
and at the W.M. Keck Observatory
which is operated as a scientific partnership among the California Institute of 
Technology, the University of California and the National Aeronautics and Space 
Administration. The Observatory was made possible by the generous financial
support of the W.M. Keck Foundation.
Keck telescope time was partially granted by NOAO,
through the Telescope System Instrumentation Program
(TSIP). TSIP is funded by NSF.}
\altaffiltext{1}{Department of Astronomy and Astrophysics, University of California Santa Cruz, Santa Cruz, CA 95064, USA}
\altaffiltext{2}{Institute of Astronomy, Madingley Road, Cambridge, UK, CB3 0HA}
\altaffiltext{3}{Institute for Astronomy, University of Hawaii, 2680 Woodlawn Drive, Honolulu, HI 96822, USA}
\altaffiltext{4}{Hubble/Morrison Fellow;~~~~email: rcooke@ucolick.org}
\date{\today}

\begin{abstract}

In this paper we analyze the kinematics, chemistry, and physical properties of a
sample of the most metal-poor damped \Lya\ systems (DLAs), 
to uncover their links to modern-day galaxies. We present evidence that
the DLA population as a whole exhibits a `knee' in the relative
abundances of the $\alpha$-capture and Fe-peak elements when the
metallicity is [Fe/H]\,$\simeq -2.0$, assuming that Zn traces the build-up of Fe-peak elements.
In this respect, the chemical evolution of DLAs 
is clearly different from that experienced by Milky Way halo stars, 
but resembles that of dwarf spheroidal galaxies 
in the Local Group. 
We also find a  close correspondence between the kinematics
of  Local Group dwarf galaxies and of high redshift 
metal-poor DLAs, which further strengthens this connection. 
On the basis of such similarities, we propose that the most
metal-poor DLAs provide us with a unique opportunity to directly 
study the dwarf galaxy population
more than ten billion years in the past, at a time when 
many dwarf galaxies were forming the bulk of their stars. 
To this end, we have
measured some of the key physical properties of the DLA gas, including
their neutral gas mass, size, kinetic temperature, density, and turbulence.
We find that metal-poor DLAs contain a warm neutral medium
with $T_{\rm gas}\simeq9600$\,K predominantly
held up by thermal pressure. Furthermore, all of the DLAs in our sample
exhibit a subsonic turbulent Mach number, implying that the gas distribution
is largely smooth. These results are among the first empirical 
descriptions of the environments where the first few generations
of stars may have formed in the Universe.
\end{abstract}


\section{Introduction: Low Metallicity Galaxies}

The population of dwarf galaxies represents the least
luminous, least massive, and perhaps the most numerous galaxies at
all redshifts in the observed Universe. However, despite their great
number, just a handful of these galaxies have been firmly identified,
owing to their intrinsically low luminosities (see \citealt{McC12} for a
continuously updated list of systems). Typically, these galaxies are
very metal-poor, and are highly varied in their physical properties
\citep{Mat98,TolHilTos09}. At present, our  understanding
of the chemical evolution and build-up of stellar mass in these galaxies
involves a careful `archaeological' study of their present-day physical
and chemical properties.

\subsection{The Nearby Universe}

For many decades, it has been appreciated that a color-magnitude
diagram (CMD) encodes the evolutionary sequence of stellar populations.
Variations in a population's age, metallicity, star formation rate and stellar initial
mass function (IMF) can in principle be recovered through a synthetic
CMD analysis \citep[e.g.][]{TolSah96}. This has
proven to be a highly successful approach to studying the star formation
history (SFH) of Local Group dwarf galaxies (see the review by
\citealt{TolHilTos09}, and references therein). 
For example, from the analysis of their CMDs, 
\citet{Wei14a} found that   
dwarf spheroidal (dSph) galaxies in the vicinity of the
Milky Way (MW) experienced highly varied
levels of star formation more than 1 Gyr ago, 
and show little to no evidence for recent star formation activity.
Beyond the Milky Way, CMD analyses for a large sample of nearby
dwarf galaxies with a range of morphological types have further
highlighted the great diversity in SFHs of low mass galaxies \citep{Wei11}.
In these systems, a considerable fraction of their stellar mass was in
place by redshift $z\sim1$, regardless of their present-day morphological
type. Furthermore, these systems appear to have experienced significant
levels of ancient star formation that took place at least 10 Gyrs ago, at
redshifts $z\gtrsim2$. Such studies have emphasized that the high redshift
Universe potentially offers a rich perspective on the evolution of
low mass galaxies.

An alternative and complementary approach to studying the evolution
of dwarf galaxies involves detailed chemical abundance
measurements of individual member stars \citep{SheBolSte98,Bon00,SheCotSar01}.
These stars condensed from gas that was enriched by the cumulative
products of previous stellar generations. The overall metallicity distribution
function therefore contains precious information on the chemical evolution
and SFH of dwarf galaxies. A timescale for the chemical enrichment can then
be teased out from the \textit{relative} metal abundances; stars of different mass
and metallicity synthesise elements in different proportions and release
their nucleosynthetic products into the surrounding interstellar medium at
different times. Therefore, deciphering the chemistry of a dwarf galaxy provides
key information on the timescales and dominant sources of enrichment. For
example, stars in the mass range $10\lesssim m_{\star} \lesssim100\,{\rm M}_{\odot}$
end their life as Type-II supernovae (SNe) on relatively short timescales ($<30$ Myr),
and are the primary source of the $\alpha$-capture elements, in addition to
some Fe-peak elements. On the other hand, Type-Ia SNe, which operate on
much longer timescales ($\sim1$ Gyr\footnote{We note, however, that the delay time distribution for
Type-Ia SNe may have a contribution from prompt Type-Ia SNe
\citep{ManDelPan06}, which operate on shorter timescales.}),
predominantly yield Fe-peak elements. Thus, one initially expects a plateau
in the ratio of $\alpha$/Fe-peak elements (due to Type-II SNe) until Type-Ia
SNe explode and contribute their Fe-peak elements, thereby lowering the $\alpha$/Fe ratio.
The downturn in the $\alpha$/Fe ratio (a.k.a. the `knee') therefore tracks the
SFH of the galaxy and its ability to retain metals
\citep[see e.g. the discussion by][]{Tin79,WheSneTru89}.

Of the known dwarf galaxies in the local Universe, only those in orbit around
the MW have resolved, bright stellar populations that allow one to measure
the metallicities of individual member stars using high resolution spectrographs,
and thereby study their chemical evolution in fine detail. In the last decade, there
have been many efforts to trace the chemical history of the brightest MW dSph
galaxies \citep{Bon04,Mon05,Sbo07,Koc08,CohHua09,CohHua10,Let10,Hen14}.
For typical MW dSphs, the knee in the $\alpha/$Fe ratio occurs at roughly 1/100 of solar metallicity
(i.e. [Fe/H] $\sim-2.0$)\footnote{Throughout this paper, we adopt the standard
notation [X/Y]$\equiv\log\,N({\rm X})/N({\rm Y}) - \log\,({\rm X}/{\rm Y})_{\odot}$,
where $N({\rm X})/N({\rm Y})$ is the number abundance ratio of element X
relative to element Y, and the ${\odot}$ symbol refers to the solar value, taken
from \citet{Asp09}.}, signaling that Type-Ia SNe contributed their enrichment at
an early stage of the dwarf galaxy's chemical evolution. This observation is very
much in agreement with inferences on the star formation history of the Milky Way dSphs from CMD
analyses --- the luminous MW dSph galaxies experienced modest levels of `bursty'
star formation over an extended period.

Deep and wide-area surveys for these low-luminosity
dSph galaxies, made possible by the Sloan Digital Sky Survey
(SDSS; \citealt{Yor00}), have uncovered a faint galaxy population --- termed
the ultra-faint dwarf (UFD) galaxies --- that extend the known suite of galaxies
to even fainter luminosities (\citealt{Wil05a,Wil05b}; an up-to-date list can be found
in \citealt{Bel13}). These `ultra-faint' galaxies appear to be an extension of the
`classical' dSph galaxy population to much lower luminosities \citep{Bel13,Wal13}.
Detailed CMD analyses of the UFDs suggest that these galaxies assembled
a significant fraction of their present-day stellar mass prior to redshift
$z\sim2$ \citep{Wei14a}. Intriguingly, there is also evidence for a
small number of carbon-enhanced metal-poor stars showing normal
(or very low) abundances of the neutron-capture elements
(i.e. the so-called `CEMP-no' stars) associated with both the
Segue 1 and Bo{\"o}tes I UFDs \citep{Nor10,Lai11,Gil13,FreSimKir14}. 
It is thought that such stars may have
condensed out of gas that was solely enriched by the first stars
\citep{Rya05,CooMad14}, making the ultra-faint dwarf galaxies the
modern-day, surviving relics of the minihalos that hosted
the very first generation of stars \citep[e.g.][]{FreBro12}.

\subsection{The High Redshift Universe}

Given the importance of dwarf galaxies in the early build-up of
cosmic structure, as well as the possibility to probe the chemistry
of the first stars and the formation of the second stellar
generation, it is of great interest to study the dwarf galaxy
population at high redshift when they were in the process of
building their stellar mass. 
However, this is a very difficult prospect in practice.
For example, the current stellar population in
Bo{\"o}tes~I \citep[at a distance of $\sim 60$\,kpc;][]{Bel06}
would correspond to a rest-frame visual
magnitude of $m_{\rm V}\simeq 41$ at redshift $z\sim 3$,
more than 10\,000 times fainter than the faintest
objects detected in the \textit{Hubble Space Telescope (HST)} 
extreme deep field \citep[][]{Ill13}.
Some assistance
can be provided by gravitational lensing \citep{Amo14,Ala14},
and indeed this is one of the primary goals of the
\textit{HST Frontier Fields}. However, such observations
are unlikely to reach the lowest luminosity dwarfs and 
will in any case be restricted to the relatively small volumes 
magnified by the gravitational lenses. 
Other `direct detection' studies have
instead searched for the putative \Lya\ emission arising from
galaxies that are experiencing very low levels of star formation
\citep{Rau08} or via fluorescent \Lya\ emission of gas-rich
galaxies that are presumably on the verge of star formation
\citep{CanLilHae12}. 
While these efforts undoubtedly contribute to our 
view of high-redshift low mass galaxies, they 
are unable to simultaneously provide detailed physical and
chemical insights into star formation in the low metallicity regime.

The  work by \citet{Wei11,Wei14a} has indicated that many
dwarf galaxies formed a considerable fraction of their present
day stellar population between redshifts $z = 1$ and 4. Therefore, at
some point in their evolution, the local dwarf galaxies must have
contained a reservoir of cold, neutral gas that allowed them to form
their small pool of stars. Such conditions are satisfied by
the generic properties of damped Lyman~$\alpha$ systems (DLAs),
absorption systems that are easily identified 
by their strong \HI\ \Lya\ absorption feature along
the line of sight to an unrelated, background light source (typically
a quasar). 
By definition, DLAs have column densities of neutral hydrogen in excess of
$N({\rm H\,\textsc{i}})\ge10^{20.3}$ \HI\ atoms ${\rm cm}^{-2}$
(\citealt{Wol86}; for a review, see \citealt{WolGawPro05}), and
are typically enriched to 1/30 of the solar metallicity at redshift $z\sim3$
\citep{Pet97,Pro03,Raf12,JorMurTho13}. This typical metallicity
is also known to evolve modestly with redshift \citep{Raf12,JorMurTho13};
an extrapolation of this relation to the present day would imply that
DLAs at redshift $z\sim0$ should be enriched to $\sim1/10$ of the
solar metallicity, similar to the metallicity of the Fornax dSph \citep[e.g.][]{Kir11a}.
Furthermore, any reasonable extrapolation of the stellar mass-metallicity
relation of DLAs proposed by \citet{Chr14} would suggest that a
typical DLA has a stellar mass consistent with a dwarf galaxy
($\sim10^{8-9}~M_{\odot}$; see also, \citealt{Mol13}).

Since the inception of DLA research,
there has been a considerable effort to image the host galaxies of
DLAs or catch them in line emission. Despite the
efforts of many, the typical DLA population has largely evaded
detection. In all cases, the imaged DLAs are either:
(1) biased toward the higher end of the DLA metallicity distribution
function and are, consequently, those most actively forming stars;
(2) undetected along the quasar sightline, but a proximate $\sim L^{\star}$
galaxy is seen at large impact parameters; or
(3) completely undetected.
Despite the difficulty of recognizing
a DLA host galaxy against the glare of the backdrop quasar, 
there are a handful of solid detections at $z\gtrsim2$
\citep{Low95,Bun99,Kul00,Chr09,Fyn10,Kro12,Kro13,Not12,Per12,JorWol14,Fum14}
which suggest that DLAs are a `mixed bag'.

Clues to the nature of DLAs from cosmological hydrodynamic
simulations have shown that the DLA cross-section
primarily comes from galaxies in halos of mass
$10^{9}-10^{11}$\,M$_{\odot}$ at $z\sim3$
\citep{Pon08,Tes09,Fum11,Cen12,van12,Bir13,RahSch14}.
Therefore, the lack of nearby $L^{\star}$ galaxies in the proximity of DLAs,
the peak of the DLA halo distribution ($\sim10^{10}$\,M$_{\odot}$),
the peak of the DLA metallicity distribution at redshift $z\simeq3$ (1/30 solar),
and the extrapolation of the DLA redshift-metallicity relation to redshift $z=0$,
all point to a \textit{typical} DLA being associated with a dwarf galaxy. However,
the measured bias factor of DLAs \citep{Coo06,Fon12,BarHae14}, under
some model assumptions, places DLAs in a typical host halo of mass
$10^{10.5}-10^{12}$\,M$_{\odot}$ at redshift $2-3$.

Several studies over the last few years have begun to focus on the
metal-poor tail of the DLA metallicity distribution function
\citep{Pet08,Pen10,Coo11a,Coo11b,Coo12,Dut14}, which currently
extends down to [Fe/H]$=-3.45$ ($z_{\rm abs}=3.6966$ toward
J0140$-$0829; \citealt{Ell10,Coo11b}).
Most of the gas in these near-pristine systems has not yet been processed
through generations of stars, and has therefore retained a primordial
composition of the light elements \citep{PetCoo12,Coo14}. Indeed, some of the
most metal-poor DLAs may have only seen the enrichment from a single
generation of Population III stars (e.g. \citealt{Coo11a}; herein, we update
the chemical composition of this DLA, see Appendix~\ref{sec:newsys}).
More generally, the most metal-poor DLAs (with metallicities [Fe/H] $\lesssim-2.0$)
exhibit an enhancement in their $\alpha$/Fe ratio, consistent with
that seen in Local Group dwarf galaxies and metal-poor
Galactic halo stars \citep{Coo11b,Raf12}.

The most metal-poor DLAs at high redshift can evidently play an
important role in furthering our understanding of
the formation of stars in extremely
low metallicity environments and at early times. Moreover, such
systems provide the only avenue 
in the foreseeable future to \textit{directly} probe the early evolution
of the smallest galaxies. In this paper, we
present measurements of the most basic physical properties of the
metal-poor DLA population (e.g. their kinetic temperature and volume
density). We also provide the first crucial links between 
high-redshift metal-poor DLAs and near-field
dwarf galaxies.

In the following section we discuss the high redshift DLA sample
used throughout this paper. To better understand the nature
of these systems, we compare the kinematics and chemistry of the
metal-poor DLA population with those of Local Group galaxies in
Section~\ref{sec:kinematics} and \ref{sec:chem} respectively.
The physical properties of a smaller sample of low metallicity DLAs
(those with the best data for the purpose) are derived in
Section~\ref{sec:properties}, giving new insights into
the physical state of the neutral gas in these systems.
In Section~\ref{sec:disc}, we discuss the nature and evolution of
the most metal-poor DLAs, before summarizing the main findings
of this work in Section~\ref{sec:conc}.

\section{The sample of metal-poor DLAs}
\label{sec:sample}

The high redshift DLA sample considered here consists of the systems
published in our recent surveys \citep{Pet08,Coo11b,Coo13,Coo14} which
focus on DLAs with iron abundance [Fe/H]~$\leq -2.0$. 
We omit two systems considered in those works
(towards the QSOs J0311$-$1722 and J1016$+$4040) where we
were unable to measure [Fe/H], but we add one new
system (in the spectrum of J1111$+$1332), and reanalyze 
another \citep[in J0035$-$0918][]{Coo11a}.
For further details on these two systems, see Appendix~\ref{sec:newsys}.
The data that are presented in this paper were all acquired with high
resolution echelle spectrographs (R~$\gtrsim30\,000$), including the
W.~M.~Keck Observatory's High Resolution Echelle Spectrograph (HIRES)
and the Very Large Telescope's Ultraviolet and Visual Echelle Spectrograph (UVES).
Relevant properties of the DLA sample considered here
are collected in Table 1. 
In Section~\ref{sec:properties}, we refine our sample selection to consider only
the DLAs where the contributions of thermal and turbulent broadening to the
absorption line profiles could be measured separately with sufficient
precision (in many cases turbulent broadening dominates over
thermal broadening, so that the latter cannot be determined independently). 
For further details, see Appendix~\ref{sec:profanal}.

\section{The kinematics of metal-poor DLAs}
\label{sec:kinematics}

The kinematic information extracted from the analysis
of  metal absorption lines in DLAs is one of the
key observables that has been used to infer the nature of the
DLA host galaxy population. The first detailed investigation
of DLA kinematics was presented by \citet{ProWol97}.
These authors introduced a test statistic, known as $v_{90}$, which
measures the velocity interval covering the innermost
90 per cent of the total optical depth of unsaturated metal
absorption. In general, DLAs exhibit a broad range
in their $v_{90}$ statistic; the distribution of values peaks near
$70$~km~s$^{-1}$ with a high velocity tail that 
extends out to $\sim 450$~km~s$^{-1}$.
Using a suite of Monte Carlo simulations, \citet{ProWol97} explored
a variety of simple galaxy models to conclude that the kinematics of the
DLA population are mostly consistent with thick, rapidly rotating disks.
An alternative possibility was proposed shortly after by \citet{HaeSteRau98}
using cosmological N-body hydrodynamic simulations. These authors
showed that the distribution of asymmetries and velocity widths of DLA
metal absorption lines can be naturally explained in the context of
hierarchical galaxy formation, where the kinematics of  DLAs
arise from the combination of ordered and random motions,
in addition to infall and merging (see, however, \citealt{ProWol10}).
Since these early works, cosmological hydrodynamic simulations
have advanced substantially, and yet they still have difficulty in 
reproducing the large velocity widths exhibited
by some DLAs \citep{Pon08,Tes09,Cen12}. 
A potential resolution to this problem
has recently been proposed by \citet{Bir15},
who interpret the largest velocity widths as arising from
the chance alignment of a DLA host galaxy and a metal-enriched Lyman Limit
System residing in a filament near the DLA \citep[see e.g.][]{Pet83}.

By combining the measured velocity widths and metallicities of
damped \Lya\ systems, several authors have proposed that DLAs
obey a `mass-metallicity' relation \citep{Led06,Mur07,Pro08,JorMurTho13,Nee13}.
Using the $v_{90}$ velocity width as a proxy for the mass of the DLA
host galaxy, DLAs with the largest velocity widths statistically show
the highest metallicities. Thus it appears that the gas distribution 
broadly traces the gravitational potential of the host galaxy.
In this picture, DLAs that exhibit the lowest metallicities should
also exhibit the most quiescent kinematics,
and thus be related to the lowest mass galaxies.

Thus, it is of interest to compare the kinematics of 
the most metal-poor DLAs to those of Local Group dwarf galaxies. 
The latter are typically characterized by the
$1\sigma$ line-of-sight dispersion of projected stellar motions or, 
if the galaxy still bears gas, from the \HI\ 21 cm line emission profile. 
In both cases, the galaxy's kinematics are quantified by
considering the 68 per cent interval of the tracer kinematics.

\begin{table}
\centering
\begin{minipage}[c]{0.5\textwidth}
    \caption{\textsc{The kinematics and chemistry of metal-poor DLAs}}
    \begin{tabular}{@{}lcccc}
    \hline
   \multicolumn{1}{c}{QSO name}
& \multicolumn{1}{c}{$z_{\rm abs}$} 
& \multicolumn{1}{c}{$v_{68}$}
& \multicolumn{1}{c}{[Fe/H]$^{\rm a}$}
& \multicolumn{1}{c}{[Si/Fe]$^{\rm a}$}\\
    \multicolumn{1}{c}{}
& \multicolumn{1}{c}{}
& \multicolumn{1}{c}{(km~s$^{-1}$)}
& \multicolumn{1}{c}{}
& \multicolumn{1}{c}{}\\
  \hline

\hline

Q0000$-$2620  &  3.390 &  12.0 &  $-2.01\pm0.09$  &  $+0.15\pm0.04$  \\
J0035$-$0918  &  2.340 &   1.6 &  $-2.94\pm0.06$  &  $+0.37\pm0.07$  \\
HS0105$+$1619 &  2.537 &   4.2 &  $-2.10\pm0.05$  &  $+0.25\pm0.06$  \\
Q0112$-$306  &  2.418 &  19.4 &  $-2.64\pm0.09$  &  $+0.25\pm0.06$  \\
J0140$-$0839  &  3.697 &  47.5 &  $-3.45\pm0.24$  &  $+0.70\pm0.21$  \\
J0307$-$4945  &  4.467  &  141  &  $-1.93\pm0.20$  &  $+0.43\pm0.18$  \\
J0831$+$3358  &  2.304 &  18.5 &  $-2.39\pm0.16$  &  $+0.38\pm0.08$  \\
Q0913$+$072   &  2.618 &  14.6 &  $-2.77\pm0.05$  &  $+0.27\pm0.02$  \\
J1001$+$0343  &  3.078 &  6.5 &  $-3.48\pm0.27$  &  $+0.62\pm0.27$  \\
J1037$+$0139  &  2.705 &   9.8 &  $-2.44\pm0.08$  &  $+0.40\pm0.04$  \\
Q1108$-$077   &  3.608 &  17.4 &  $-1.96\pm0.07$  &  $+0.42\pm0.03$  \\
J1111$+$1332  &  2.271 &   3.0 &  $-2.27\pm0.04$  &  $+0.32\pm0.02$  \\
J1337$+$3152  &  3.168 &  18.5 &  $-2.74\pm0.30$  &  $+0.06\pm0.26$  \\
J1340$+$1106  &  2.508 &  24.6 &  $-2.04\pm0.06$  &  $+0.18\pm0.03$  \\
J1340$+$1106  &  2.796 &  10.4 &  $-2.15\pm0.06$  &  $+0.32\pm0.03$  \\
J1358$+$6522  &  3.067 &   3.4 &  $-2.60\pm0.07$  &  $+0.31\pm0.03$  \\
J1419$+$0829  &  3.050 &  17.3 &  $-2.27\pm0.11$  &  $+0.24\pm0.02$  \\
J1558$-$0031  &  2.702 &  12.4 &  $-2.03\pm0.04$  &  $+0.26\pm0.05$  \\
J1558$+$4053  &  2.554 &   8.2 &  $-2.70\pm0.07$  &  $+0.21\pm0.07$  \\
Q1946$+$7658  &  2.844 &   7.0 &  $-2.50\pm0.06$  &  $+0.32\pm0.02$  \\
Q2059$-$360    &  3.083  &  32.7  &  $-1.97\pm0.08$  &  $+0.34\pm0.06$ \\
J2155$+$1358  &  4.212 &  26.4 &  $-2.15\pm0.25$  &  $+0.28\pm0.23$  \\
Q2206$-$199   &  2.076 &   7.5 &  $-2.55\pm0.04$  &  $+0.30\pm0.02$  \\

  \hline
    \end{tabular}

    \smallskip
    
\hspace{0.5cm}$^{\rm a}${We have adopted solar abundances of $12+\log$(Si/H)$_{\odot}=7.51$
and $12+\log$(Fe/H)$_{\odot}=7.47$ from \citet{Asp09}.}\\

    \label{tab:kinsife}
\end{minipage}
\end{table}

Therefore, to facilitate a comparison between $z\sim3$ DLAs and
Local Group galaxies, we define a new velocity width statistic
for DLAs, hereafter denoted $v_{68}$, which is qualitatively similar
to the $v_{90}$ statistic proposed by \citet{ProWol97}. Specifically, $v_{68}$
is the velocity interval containing the innermost 68 per cent of the total
optical depth. Although our $v_{68}$ statistic cannot be directly related
to a \textit{stellar} velocity dispersion, as measured in the Local Group
dwarf galaxies, it provides a simple conversion to the velocity dispersion
of the gas. Furthermore, compared with the $v_{90}$ statistic, $v_{68}$
is much less sensitive to weak absorption that is well separated
from the dominant optical depth. In order to measure $v_{68}$
from the observed profiles of the absorption lines in a DLA,
it is necessary to correct for two sources of broadening
that are unrelated to the bulk motions of the gas,
namely thermal (i.e. microscopic)
and instrumental (i.e. due to the spectrograph with which the 
data were recorded) 
broadening. In the most
quiescent DLAs, where the absorption is confined to a
single velocity component,
the correction is straightforward (if the
thermal and instrumental broadening are known).
In this case, $v_{68}\equiv2\sigma_{\rm los}=\sqrt{2}\,b_{\rm turb}$.

While it is generally the case that the velocity structure of 
DLAs with [Fe/H]\,$\leq -2.0$ is relatively simple, 
in many of the DLAs in Table~\ref{tab:kinsife}
the absorption is spread over a few velocity components. 
In such cases, we
employed the following procedure to determine $v_{68}$.
First, we generated synthetic \SiII\ (or \FeII) line profiles for each
system using only the column densities and turbulent broadening
derived from the cloud model that best fits the absorption
lines as described in Appendix A (i.e. we did not include
thermal and instrumental broadening for the synthetic profiles). 
\SiII\ and \FeII\ are
the optimum species to use for measuring the velocity fields
in these DLAs, since there are generally many \SiII\ and \FeII\ lines of
widely different oscillator strengths, allowing the optical depth
to be determined for the full range of velocities
over which absorption takes place. Furthermore, of
the lines that are often seen in metal-poor DLAs, \SiII\ and \FeII\
are the least affected by thermal broadening\footnote{For systems
where $v_{68}$ was at least twice the instrumental FWHM, we found
that $v_{68}$ can accurately be measured directly from the data,
without the need to generate synthetic profiles that ignore the thermal
and instrumental contributions to the line broadening.}. We then
converted the synthetic line profiles to an optical depth profile,
and calculated the velocity interval containing the innermost
68 per cent of the total integrated optical depth. Finally, we relate our
velocity statistic to the $1\sigma$ line-of-sight velocity dispersion by
the relation $\sigma_{\rm los}=v_{68}/2$. We note that our analysis
only considers the velocity width traced by the low ion metal absorption
lines (i.e. C$^{+}$, O$^{0}$, Si$^{+}$, Fe$^{+}$), which are known
to predominantly trace the \HI\ gas. 

\begin{figure*}
  \centering
  \hspace{-0.75cm}\includegraphics[angle=0,width=14.5cm]{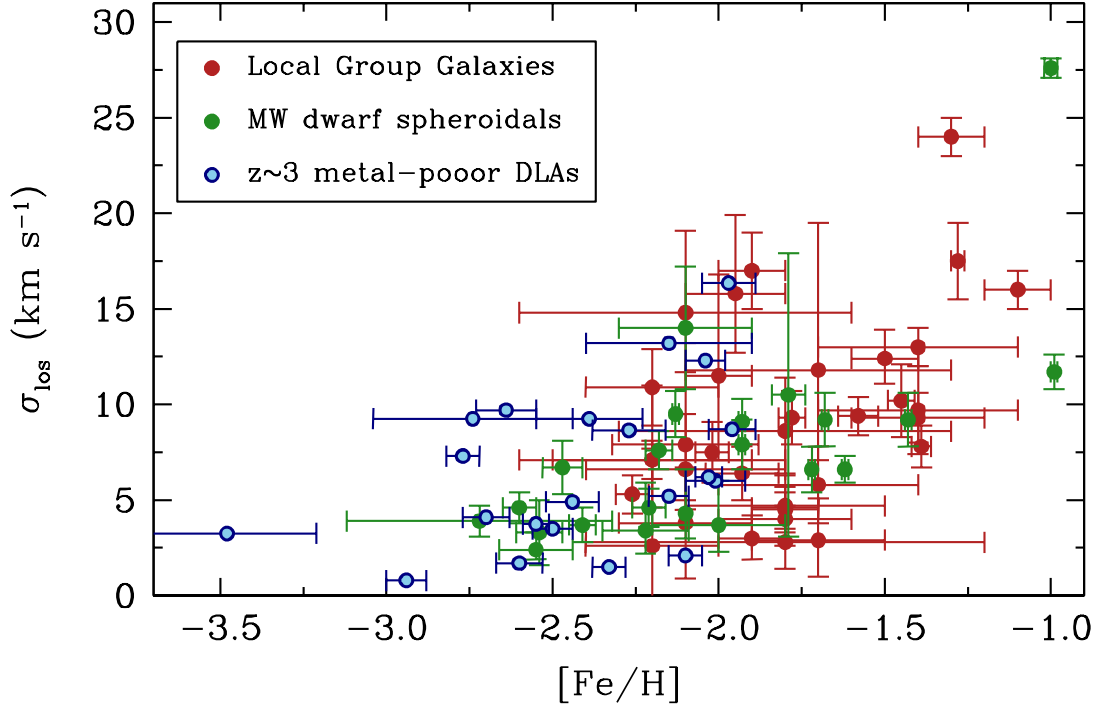}
  \caption{The line-of-sight stellar velocity dispersions and
metallicities for a sample of nearby low-metallicity galaxies
(red and green symbols; \citealt{McC12}) are compared to the
most metal-poor DLAs at $z\simeq3$ (blue symbols). The green
symbols highlight the Milky Way dwarf spheroidal galaxies.
  }
  \label{fig:sigMH}
\end{figure*}

\begin{figure*}
  \centering
  \hspace{-0.75cm}\includegraphics[angle=0,width=14.5cm]{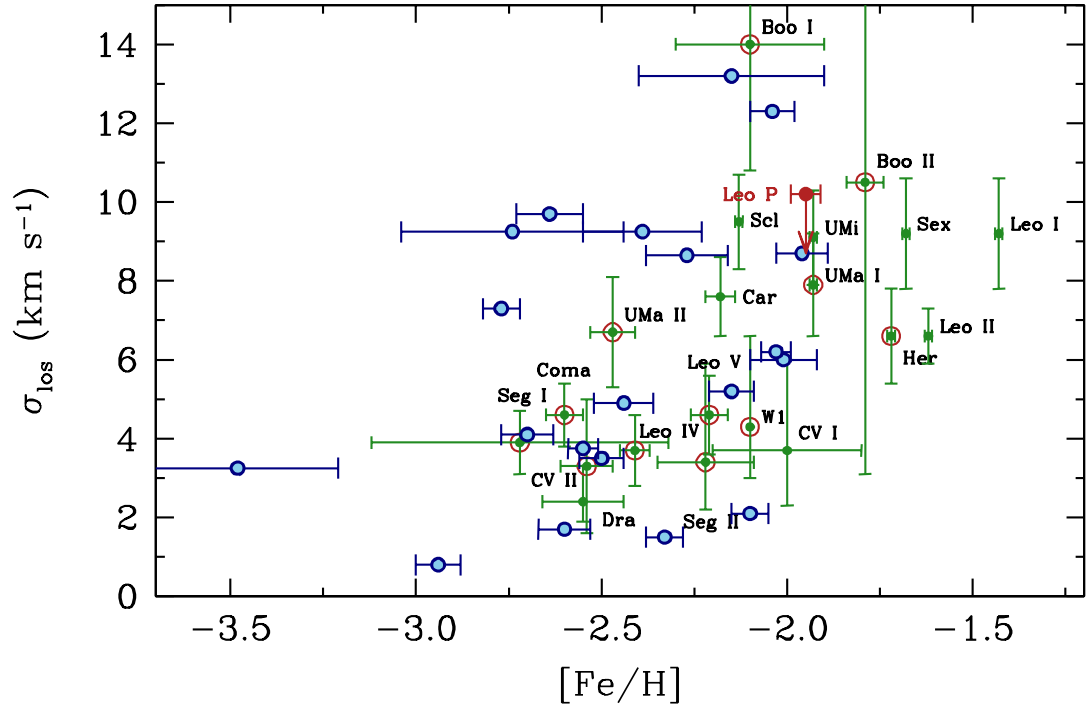}
  \caption{A zoomed in version of Fig.~\ref{fig:sigMH}, comparing only
the Milky Way dwarf spheroidals (green symbols; labelled) and the most
metal-poor DLAs (blue symbols). The green symbols with red circles
correspond to the ultra-faint dwarf galaxies, with total luminosities
$\le10^{5}\,L_{\odot}$. We also show the recently discovered nearby
gas-rich dwarf galaxy Leo P, where the velocity dispersion is measured
from \HI\ 21 cm observations and the Fe abundance is derived from the
associated \HII\ region, assuming [Fe/H] = [O/H] $- 0.4$ (red dot;
\citealt{Gio13}, \citealt{Ski13}). We consider the line-of-sight velocity
dispersion for Leo P as an upper limit, since thermal broadening
of the line profile may contribute  to the line width.
  }
  \label{fig:sigMHzoom}
\end{figure*}

The full sample of DLAs
for which we have measured $v_{68}$ is collected in Table~\ref{tab:kinsife},
which also includes values of [Fe/H] `metallicity', and of 
[Si/Fe], representative of the ratio of $\alpha$-capture to Fe-peak elements.
The data in the last two columns were culled 
from the references  given in Section~2.
In Figure~\ref{fig:sigMH}, we compare the measured kinematics of the
DLAs in our sample to those of Local Group galaxies as compiled by
\citet{McC12}. To allow for a closer comparison to the MW
dSph population, we also present a zoom-in version of
this comparison in Figure~\ref{fig:sigMHzoom}, where we have
placed a red circle around the galaxies that have been discovered
since the start of the SDSS survey (i.e. those classified as UFDs).
There is evidently a very close correspondence between the kinematics
and metallicity for the galaxies displayed in Figures~\ref{fig:sigMH}
and \ref{fig:sigMHzoom} despite the diversity of techniques
employed to measure these two quantities.

For a given DLA, we are unable to make a direct connection between our
$v_{68}$ statistic and the DLA host potential; since we are unaware of the 
geometry of the DLA host galaxy, this connection may depend on
the inclination of the DLA relative to the sightline probed by the quasar.
However, for a sample of systems \textit{at a given metallicity},
each DLA will be intersected at a random inclination angle, so that
statistically we may be justified in considering the range of velocity 
widths exhibited by DLAs at a given metallicity. 
From inspection of Figures~\ref{fig:sigMH} and
\ref{fig:sigMHzoom}, it can be seen that systems 
with higher velocity widths tend to have higher
metallicities. Similarly, the lowest velocity width systems are associated
with the lowest metallicity DLAs. We therefore conclude that the most
metal-poor DLAs are generally associated with the lowest mass galaxies
that can potentially still form stars at $z\sim3$.

A clear outlier in the velocity width-metallicity relation for metal-poor
DLAs is the DLA towards J0140$-$0839 (\citealt{Ell10}; included in
Table~\ref{tab:kinsife} but not shown in Figures~\ref{fig:sigMH} and \ref{fig:sigMHzoom}
as it is off-scale in those plots).
This DLA has one of the highest velocity width measurements of our sample and one of
the lowest metallicities. This example highlights the main caveat
of our technique; with quasar absorption line spectroscopy, we
are unable to distinguish between a DLA in a single halo with a
broad velocity distribution or two individual DLAs in close proximity
with a small relative velocity along the line of sight. Given that this
system exhibits two distinct absorption
components separated by $\sim$45\,km~s$^{-1}$,
each with $v_{68}\simeq6-10$~km~s$^{-1}$,
we speculate that its abnormally high velocity width
may result from a chance alignment or interaction
of two very metal-poor DLAs (which, coincidentally,
are also close to the QSO redshift).

Clearly, a \textit{direct} comparison between the kinematics
of redshift $z\sim3$ metal-poor DLAs and  Local Group dwarf galaxies is
fraught with uncertainty, given the different physics that operates
on gas versus stars. Nevertheless, on the basis of this comparison, we
conclude that the general trend and range of velocity widths
exhibited by the most metal-poor DLAs are similar
to those seen in nearby dwarf galaxies.

\section{The chemistry of metal-poor DLAs}
\label{sec:chem}

The chemical composition and relative element abundances of DLAs
provide crucial information on the chemical evolution, star formation history,
and ultimately the nature of the host galaxies that give rise to the absorption.
In this section, we present the current complement of data
that allows us to infer the most likely star formation history of typical DLAs,
as traced by the relative abundance of $\alpha$-capture and Fe-peak
elements. At metallicities less than $\sim 1/100$ of solar, 
the production of $\alpha$ elements 
in DLAs is best traced by considering
the \SiII\ ion, which has numerous transitions with a variety of 
oscillator strengths and over a range of wavelengths that are redshifted
into a convenient portion of the visible spectrum at redshifts
$z = 2$--3. For the same reasons, the most reliable
tracer of the Fe-peak elements at such metallicities is \FeII.

\begin{figure*}
  \centering
  \hspace{-1cm}\includegraphics[angle=0,width=14.0cm]{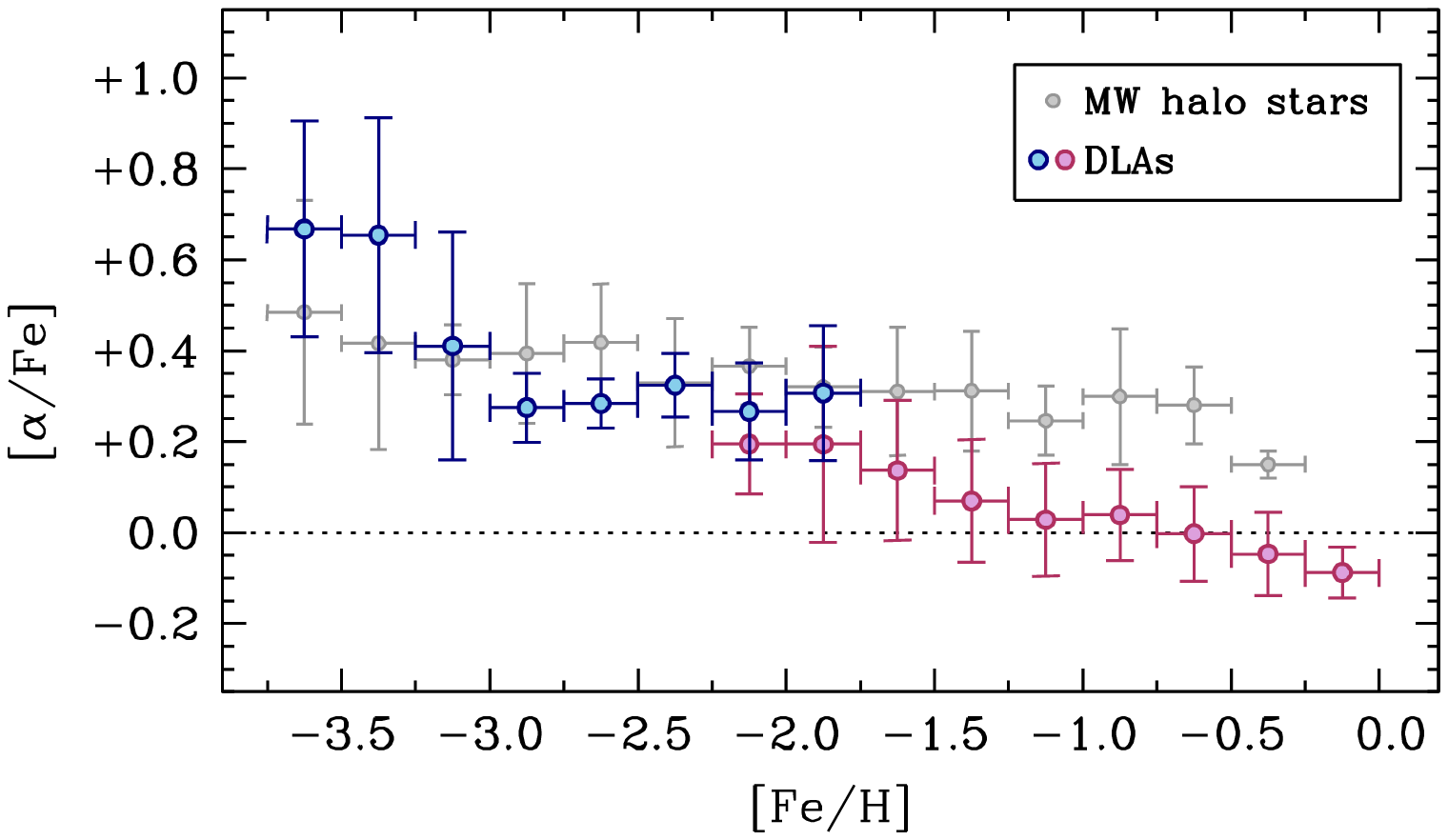}
  \caption{The [$\alpha$/Fe] ratio in DLAs (red symbols denote [S/Zn] and [Zn/H] from
  \citealt{Vla11}; blue symbols are for [Si/Fe] and [Fe/H]
  from our work) are compared to the typical [$\alpha/{\rm Fe}$] measured in
  Milky Way halo stars (gray symbols; \citealt{Ven04}). For every 0.25~dex metallicity bin,
  we plot the mean value of the population, weighted by the errors, and represent the $1\sigma$
  dispersion in the population with error bars.
  The horizontal dotted line is drawn at the solar value of the $\alpha$/Fe ratio.
  Note that the evolution of [$\alpha$/Fe] for DLAs would be slightly different if
  [Zn/Fe]$~\neq 0.0$ when [Fe/H]$~>-2.0$ (see text for the relevant discussion).
  }
  \label{fig:aFeplot}
\end{figure*}

For DLAs with metallicities greater than $\sim1/100$ of  solar,
some $\alpha$-capture and Fe-peak elements may suffer from
depletion onto dust grains \citep{Pet97,Ake05,Vla11}
so that the [Si/Fe] abundance measured in the gas phase
may be different from the intrinsic [$\alpha$/Fe] abundance set by chemical
evolution. To avoid the effects of dust depletion, S and Zn are often considered
the most appropriate combination to trace the relative abundance of $\alpha$/Fe
in the interstellar medium;
both S and Zn have very little affinity for dust grains and, in any case, have
an almost identical condensation temperature. 
Moreover, both \SII\ and \ZnII\
offer multiple transitions that are usually straightforward to measure and
rarely suffer from saturation.\footnote{It is well-established that in DLAs
the first ions are the dominant stages of the corresponding elements. Notable 
exceptions are N and O which are predominantly neutral
\citep[see][and section 5.2 below]{Mor73}.}

However, with current telescope facilities, the strongest \ZnII\ absorption lines
become too weak to be reliably measured when the DLA metallicity is
$\lesssim 1/100$ of solar. Therefore, in order to track the full behaviour
of the $\alpha$/Fe ratio in DLAs, we must combine measurements of [Si/Fe]
at low metallicity with [S/Zn] measures at high metallicity. 
While using different element ratios in different metallicity
regimes is clearly not ideal, we believe that systematic offsets 
between these two pairs of elements are likely to be small.
Combining measurements of Si and S seems justified, since detailed nucleosynthesis
calculations of massive stars have illustrated that Si and S are produced
in solar relative proportions over a wide range of metallicity and stellar mass
\citep{WooWea95,ChiLim04}. Furthermore, measurements of Zn and Fe
in Galactic halo stars with [Fe/H]\,$\gtrsim-2.0$ have shown that Zn/Fe
accurately traces the solar ratio \citep{Sai09}. For these reasons, we
propose that the combination of [Si/Fe] at the lowest metallicities and
[S/Zn] at the highest metallicities provides the most reliable diagnostic
of the chemical evolution of $\alpha$-capture and Fe-peak elements
in DLAs, when drawing comparisons with Galactic halo stars.

Figure~\ref{fig:aFeplot} shows the distribution of
[$\alpha$/Fe] values in DLAs in 0.25 dex wide bins
of `metallicity'  over nearly four orders of magnitude
in [Fe/H], from [Fe/H]\,$= -3.75$ to $0.0$, together
with the standard deviation appropriate to each bin
(represented by the error bars). To account for
the uncertainty in both [$\alpha$/Fe] and [Fe/H],
we generated 1000 Monte Carlo realizations of
each measurement prior to binning. Therefore, there
is a small amount of correlation between neighboring
symbols. Red symbols denote values of [S/Zn] and [Zn/H] 
(as proxies for [$\alpha$/Fe] and [Fe/H] respectively)
from the compilation of such measurements 
by \citet{Vla11}, while the blue circles correspond 
to the values of [Si/Fe] and [Fe/H] collected in Table~\ref{tab:kinsife}.
Both sets of data are constructed exclusively from measurements using 
spectra of high S/N ratio and resolution. 
Note the good agreement between red and blue symbols
near [Fe/H]\,$\simeq -2.0$, where the two data sets
overlap.

Also shown in Figure~\ref{fig:aFeplot} are equivalent data
for Galactic halo stars (gray circles; \citealt{Ven04}). 
In this case, the [$\alpha$/Fe] values were calculated
from the average of the ratios Mg/Fe, Si/Fe, and Ca/Fe,
depending on the data available for a particular star.
This is in line with many previous stellar studies, and
is supported by the results of nucleosynthesis
calculations showing that the relative yields of 
Mg, Si, S, and Ca from massive stars are in
solar relative proportions when integrated over a Salpeter-like stellar
initial mass function \citep{WooWea95,ChiLim04}. 

In order to demonstrate that a typical DLA experiences a different
star formation history to the galaxies that built the Milky Way stellar halo, we
perform a proof by contradiction, with the starting hypothesis
that a typical DLA experiences an identical SFH
as the galaxies that predominantly built
the stellar halo of the Milky Way. Under this hypothesis, we
are justified in our assumption that [S/Si]~$=0.0$ and
[Zn/Fe]~$=0.0$ for [Fe/H]~$\gtrsim-2.5$.
As can be appreciated from the comparison presented in
Figure~\ref{fig:aFeplot}, halo stars and DLAs agree in exhibiting
an [$\alpha$/Fe]~$\simeq+0.3$ plateau when [Fe/H]~$\lesssim-2.0$,
as noted previously by \citet{Raf12}.  A similar plateau
is also observed in oxygen:  [O/Fe]~$\simeq+0.4$ 
in DLAs with  [Fe/H]~$\lesssim-2.0$ \citep{Coo11b}.
These plateaux represent the enhanced [$\alpha$/Fe] level set
by massive stars\footnote{The different levels of these plateaux
at low metallicities, corresponding to [O/Si]~$\simeq+0.1$, is
in good agreement with low metallicity nucleosynthesis calculations
\citep{WooWea95,ChiLim04,HegWoo10}.}. However, for metallicities
[Fe/H]~$\gtrsim-2.0$, the DLA population exhibits a gentle decline in
[$\alpha$/Fe] towards solar values, presumably marking an increased
contribution of Fe-peak elements from Type Ia SNe. 
In this aspect, DLAs are clearly different from Galactic halo stars
which \textit{maintain} an approximately constant $\alpha$-enhancement 
from [Fe/H]\,$= -3.75$ to [Fe/H]\,$\sim - 0.75$. This disagreement
contradicts our initial hypothesis, therefore implying that a
\textit{typical} DLA experienced a different chemical enrichment
history from that of the Milky Way stellar halo.

\begin{figure*}
  \centering
  \includegraphics[angle=0,width=15.0cm]{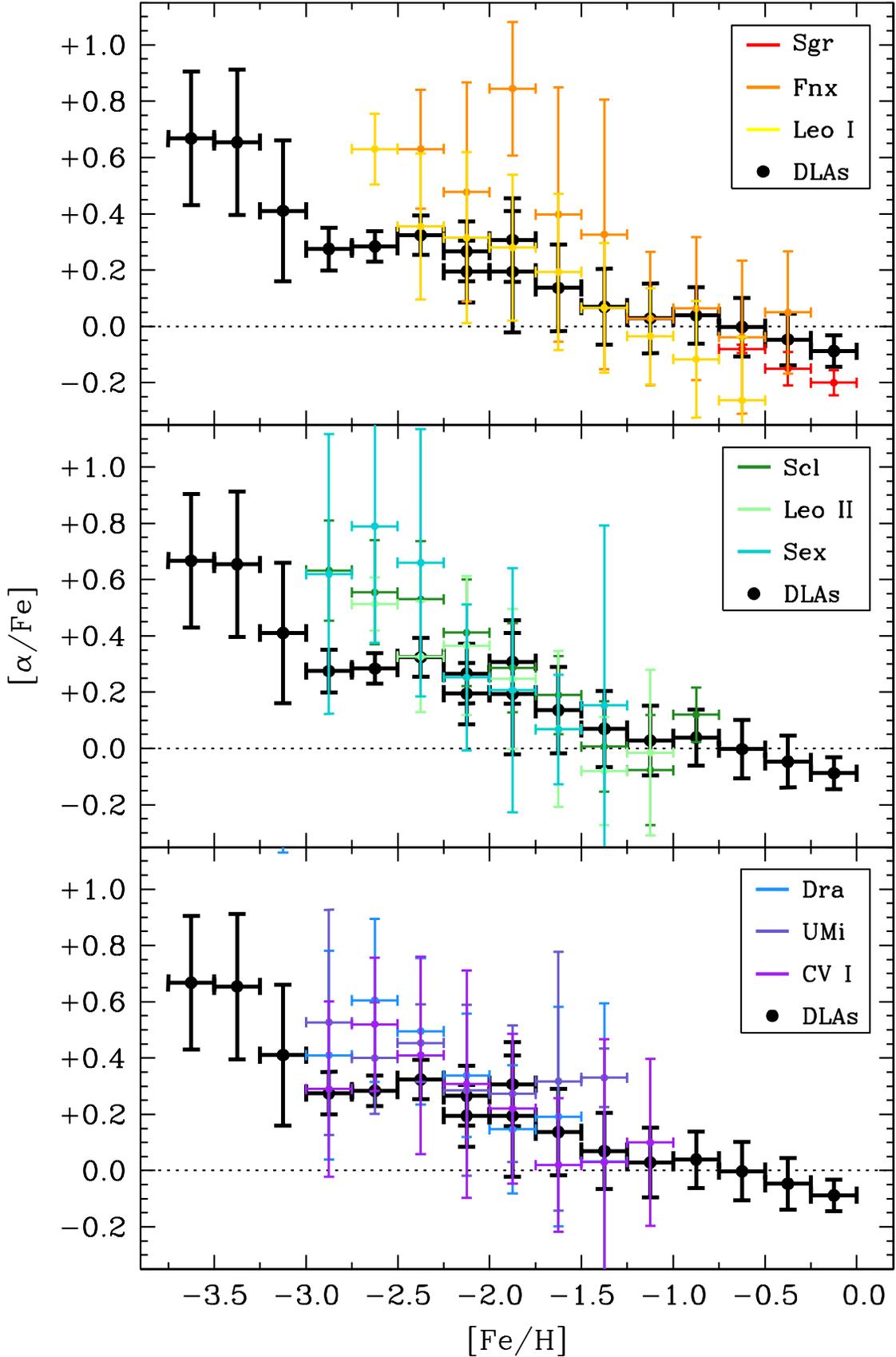}
  \caption{The [$\alpha$/Fe] ratio in DLAs (black symbols; cf. Figure~\ref{fig:aFeplot}) are
  compared to the typical [$\alpha/{\rm Fe}$] measured in the classical
  Milky Way satellite dwarf galaxies. For clarity, the dSph galaxies are color-coded by
  their V-band absolute magnitude, where red represents the brightest dSph (Sgr) and
  purple represents the faintest dSph (CV I). The data for
  Sgr are taken from \citet{Bon04}, \citet{Mon05}, and \citet{Sbo07}. The data for the
  remaining dSphs are from \citet{Kir11b}.
  }
  \label{fig:aFedSph}
\end{figure*}

Rather, the dependence of the $\alpha$/Fe ratio on metallicity in
high-redshift DLAs seems to be more akin to that seen in
local dSph galaxies \citep{Kir11b}, as can be 
realised by considering Figure~\ref{fig:aFedSph}.
For clarity, we have divided the dSph into three subsets,
as indicated; in each of the three panels we show the same set of
DLA measurements as in Figure~\ref{fig:aFeplot}, now 
uniformly indicated by black symbols, assuming that
[Zn/Fe]~$=0$. In general, there appears to be much in
common between the chemical evolutions
of MW dSph galaxies on the one hand and
of DLAs at redshift $z_{\rm abs}~\sim~2-3$ on the other
(we investigate our assumption that [Zn/Fe]~$=0$ for dSph
stars directly below).
We also note a similarity in the DLA measures of [$\alpha$/Fe]
and [Fe/H] to those recently determined in the dwarf galaxies
associated with M31, in particular AndV \citep[see Figure~4 of][]{Var14}.

Regarding the MW systems, there may be some tension towards the
lowest metallicities, where dSph stars are somewhat
more $\alpha$-enhanced relative to the DLAs. 
This could
be the result of small number statistics, modelling techniques,
or genuine differences between DLAs and the dSphs. 
We note
that several studies working with many fewer stars than \citet{Kir11b},
but with somewhat higher S/N ratio and resolution, have reported
an [$\alpha$/Fe] plateau for low metallicity stars in some dSph
galaxies at a level that is consistent with that seen in DLAs.
Some examples include Fornax \citep{Hen14},
Sculptor (\citealt{Sta13}; \citealt{Hil12}),
and Ursa Minor \citep{CohHua10}.
We also note that our comparison between DLAs and
dSph galaxies for metallicities [Fe/H]~$\gtrsim-2.0$ assumes
that S traces Si whilst Zn follows Fe. As argued earlier,
S and Si are believed to be produced in solar
relative proportions by massive stars, independent of chemical
evolution. However, Zn may not always follow Fe in the dSph galaxies;
unlike the stars in the Milky Way disc and halo \citep{Sai09}, the
small number of measurements of [Zn/Fe] available in a few dSph
galaxies suggest that [Zn/Fe] might vary with metallicity. Specifically,
[Zn/Fe]~$=0.0$ near [Fe/H]~$=-2.0$ for the dSph galaxies,
and gradually decreases to [Zn/Fe]~$\simeq-0.2$ to $-0.5$
near [Fe/H]~$=-1.0$
\citep{SheCotSar01,Sbo07,CohHua09,CohHua10}.
This difference between the dSph galaxies and the Milky Way
halo stars is likely due to their different star formation history.\footnote{We also
note that [Zn/Fe] is typically solar or subsolar when [Fe/H]~$>-2.0$ in every galaxy observed
locally, all of which exhibit a range of star formation histories. The supersolar
[Zn/Fe] abundance that is commonly observed in DLAs is due to Fe being
depleted onto dust grains (see e.g. \citealt{Raf12}).}
Therefore, to enable a fair comparison between DLAs
and dwarf galaxies one should shift the DLA points to
somewhat lower [$\alpha$/Fe] and higher [Fe/H] values
when [Fe/H]~$\gtrsim-2.0$.
Thus, if the chemical evolution of a typical DLA is similar to
that of a dwarf galaxy, such that [Zn/Fe] becomes subsolar for
[Fe/H]~$\gtrsim-2.0$, the DLA `knee' will become even more
pronounced.

With these reservations in mind, the similarity in the behaviour 
of [$\alpha$/Fe] as a function of [Fe/H]  in DLAs and dSph galaxies 
displayed in Figure~\ref{fig:aFedSph} in the metallicity range $-2.0\leq$~[Fe/H]~$\leq-1.0$
suggests that the chemical evolution of a typical DLA ([Fe/H]~$\simeq-1.5$)
is more akin to that of a dwarf galaxy than the Milky Way. This similarity
between the [$\alpha$/Fe] abundance of DLAs and the
MW dSph galaxies was also pointed out by \citet{Bon04},
albeit with many fewer dSph stars and DLAs than in the present study.
Our combined sample of 54 DLAs with reliable [$\alpha$/Fe]
measurements covering over 3 decades in metallicity has allowed
us to trace the all-important `knee' in the [$\alpha/$Fe] ratio.
Of course, it is difficult to draw firm conclusions about the entire
DLA population; for example, the high-metallicity DLAs considered
in Fig.~\ref{fig:aFedSph} are unlikely to evolve as dwarf galaxies.
Nevertheless, these data support a picture where a \textit{typical}
DLA likely traces the chemical evolution of a dwarf galaxy. Therefore,
by extension, the most metal-poor DLAs likely represent the least
chemically evolved dwarf galaxies at redshifts $z\sim3$ that still
have the potential to form stars. Future, more extensive measurements
of S and Zn in dSph stars and low metallicity DLAs would allow us
to test this picture further.
\begin{table*}
\centering
\begin{minipage}[c]{0.75\textwidth}
    \caption{\textsc{Measured physical properties of metal-poor DLAs$^{\rm a}$}}
    \begin{tabular}{@{}lccccccccc}
    \hline
   \multicolumn{1}{c}{\multirow{2}{*}{QSO name}}
& \multicolumn{1}{c}{\multirow{2}{*}{$z_{\rm em}$}} 
& \multicolumn{1}{c}{\multirow{2}{*}{$z_{\rm abs}$}} 
& \multicolumn{1}{c}{$b_{\rm turb}$}
& \multicolumn{1}{c}{$T_{\rm gas}\,^{\rm b}$}
& \multicolumn{1}{c}{\multirow{2}{*}{$\log \frac{\textrm{{\it N}(H\,\textsc{i})}}{\textrm{cm$^{-2}$}}^{\rm c}$}}
& \multicolumn{1}{c}{\multirow{2}{*}{$\log \frac{\textrm{{\it N}(H\,\textsc{i})}}{\textrm{cm$^{-2}$}}^{\rm d}$}}
& \multicolumn{1}{c}{\multirow{2}{*}{[O/H]$^{\rm e}$}}
& \multicolumn{1}{c}{\multirow{2}{*}{[Fe/H]$^{\rm e}$}}
& \multicolumn{1}{c}{\multirow{2}{*}{[Si/Fe]$^{\rm e}$}}\\
    
& 
& 
& \multicolumn{1}{c}{(km~s$^{-1}$)}
& \multicolumn{1}{c}{(K)}
&
&
& 
& 
& \\
  \hline

\multirow{2}{*}{J0035$-$0918}   &  \multirow{2}{*}{2.420}  &     $2.340097$  &     $1.1$  &     $7800$   &    $20.43$  &    $20.43$  &    $-2.44$  &    $-2.94$  &    $+0.37$  \\
                                &                          &  $\pm0.000001$  &  $\pm0.4$  &  $\pm1600$   &  $\pm0.04$  &  $\pm0.04$  &  $\pm0.07$  &  $\pm0.06$  &  $\pm0.07$   \\
\hline

\multirow{2}{*}{HS\,0105$+$1619}  &  \multirow{2}{*}{2.652}  &     $2.536509$  &     $3.0$  &    $10800$  &    $19.415$  &    $19.426$  &    $-1.76$  &    $-2.10$  &    $+0.25$  \\
                                  &                          &  $\pm0.000001$  &  $\pm0.1$  &   $\pm200$  &  $\pm0.007$ &  $\pm0.006$  &  $\pm0.02$  &  $\pm0.05$  &  $\pm0.06$  \\
\hline

Q0913$+$072         &  \multirow{4}{*}{2.785}  &     $2.618435$  &     $4.5$  &     $8400$    &    $20.09$  &  &   $-2.44$  &    $-2.73$  &    $+0.28$  \\
(comp. A)           &                          &  $\pm0.000001$  &  $\pm0.2$  &  $\pm1300$       &  $\pm0.02$  & 20.312 &  $\pm0.03$  &  $\pm0.02$  &  $\pm0.02$  \\
Q0913$+$072         &                          &     $2.618289$  &     $3.9$  &     $7700$    &    $19.84$  & $\pm0.008$ &    $-2.35$  &    $-2.85$  &    $+0.29$  \\
(comp. B)           &                          &  $\pm0.000001$  &  $\pm0.2$  &   $\pm200$    &  $\pm0.03$  & &  $\pm0.02$  &  $\pm0.04$  &  $\pm0.02$  \\
\hline

\multirow{2}{*}{J1001$+$0343}  &  \multirow{2}{*}{3.198}  &     $3.078404$  &     $4.6$  &    $17000$  &      $20.21$  & $20.21$  &    $-2.64$  &    $-3.48$  &    $+0.62$  \\
                               &                          &  $\pm0.000002$  &  $\pm0.7$  &  $\pm8000$  &  $\pm0.05$  & $\pm0.05$  &  $\pm0.06$  &  $\pm0.27$  &  $\pm0.27$  \\
\hline

\multirow{2}{*}{J1111$+$1332}  &  \multirow{2}{*}{2.420}  &     $2.270940$  &     $2.2$  &    $13000$  &    $20.39$  &    $20.39$  &    $-1.92$  &    $-2.27$  &    $+0.32$  \\
                               &                          &  $\pm0.000001$  &  $\pm0.1$  &  $\pm2000$  &  $\pm0.04$  & $\pm0.04$  &  $\pm0.08$  &  $\pm0.04$  &  $\pm0.02$  \\
\hline

\multirow{2}{*}{J1358$+$6522}  &  \multirow{2}{*}{3.173}  &     $3.0672594$  &     $2.4$  &     $5600$  &    $20.22$  &    $20.495$  &    $-2.06$  &    $-2.60$  &    $+0.31$  \\
                               &                          &  $\pm0.0000005$  &  $\pm0.5$  &  $\pm1100$  &  $\pm0.06$  & $\pm0.008$  &  $\pm0.06$  &  $\pm0.07$  &  $\pm0.03$  \\
\hline

J1419$+$0829        &  \multirow{4}{*}{3.030}  &     $3.049654$  &     $2.8$  &     $9700$    &    $19.93$   & &   $-1.82$  &    $-2.17$  &    $+0.23$  \\
(comp. A)           &                          &  $\pm0.000001$  &  $\pm0.2$  &   $\pm100$  &  $\pm0.02$   & 20.392 & $\pm0.03$  &  $\pm0.04$  &  $\pm0.03$  \\
J1419$+$0829        &                          &     $3.049843$  &     $5.3$  &    $11600$  &      $20.209$  &$\pm0.003$ &   $-1.99$  &    $-2.37$  &    $+0.24$  \\
(comp. B)           &                          &  $\pm0.000001$  &  $\pm0.2$  &   $\pm200$  &   $\pm0.008$  & & $\pm0.01$  &  $\pm0.03$  &  $\pm0.03$  \\
\hline

J1558$-$0031        &  \multirow{4}{*}{2.823}  &     $2.702318$  &     $3.7$  &     $7200$    &    $20.27$ &   &    $-1.52$  &    $-1.90$  &    $+0.22$  \\
(comp. A)           &                          &  $\pm0.000003$  &  $\pm0.2$  &  $\pm1000$  &    $\pm0.08$  &20.75 & $\pm0.09$  &  $\pm0.08$  &  $\pm0.04$  \\
J1558$-$0031        &                          &     $2.702422$  &     $1.5$  &     $9300$  &       $20.29$  & $\pm0.03$ &   $-1.47$  &    $-1.81$  &    $+0.28$  \\
(comp. B)           &                          &  $\pm0.000002$  &  $\pm0.3$  &   $\pm900$  &    $\pm0.11$  & &  $\pm0.12$  &  $\pm0.11$  &  $\pm0.07$  \\
\hline

\multirow{2}{*}{Q2206$-$199$^{\rm f}$}  &  \multirow{2}{*}{2.559}  &     $2.076229$  &     $5.3$  &    $12200$  &    $20.43$  &$20.43$  &    $-2.04$  &    $-2.55$  &    $+0.30$  \\
                               &                          &  $\pm0.000001$  &  $\pm0.2$  &  $\pm3000$  &  $\pm0.04$  &   $\pm0.04$  &  $\pm0.06$  &  $\pm0.04$  &  $\pm0.02$  \\
\hline

  \hline
    \end{tabular}

    \smallskip

\hspace{0.01cm}$^{\rm a}${The DLAs in this table are a subset of our full sample (see Section~\ref{sec:properties}).}

    \smallskip

\hspace{0.01cm}$^{\rm b}${Our technique implicitly assumes that the gas is distributed according to a Maxwell-Boltzmann
distribution. The uncertainty of the kinetic temperature measurements reported here only reflect the random uncertainty
based on the statistical fluctuations in the data.}
    
    \smallskip

\hspace{0.01cm}$^{\rm c}${\HI\ column density for a given absorption component of the DLA.}
    
    \smallskip

\hspace{0.01cm}$^{\rm d}${Total \HI\ column density of the DLA.}
    
    \smallskip

\hspace{0.01cm}$^{\rm e}${We have assumed a solar abundance of 12+$\log$(O/H)$_{\odot}=8.69$, 12+$\log$(Si/H)$_{\odot}=7.51$, 12+$\log$(Fe/H)$_{\odot}=7.47$ \citep{Asp09}.}
    
    \smallskip

\hspace{0.01cm}$^{\rm f}${Values are derived from \citet{Pet08} and \citet{Car12}.}\\
    
    \label{tab:mphysprop}
\end{minipage}
\end{table*}

\begin{figure*}
  \centering
  \includegraphics[angle=0,width=8.5cm]{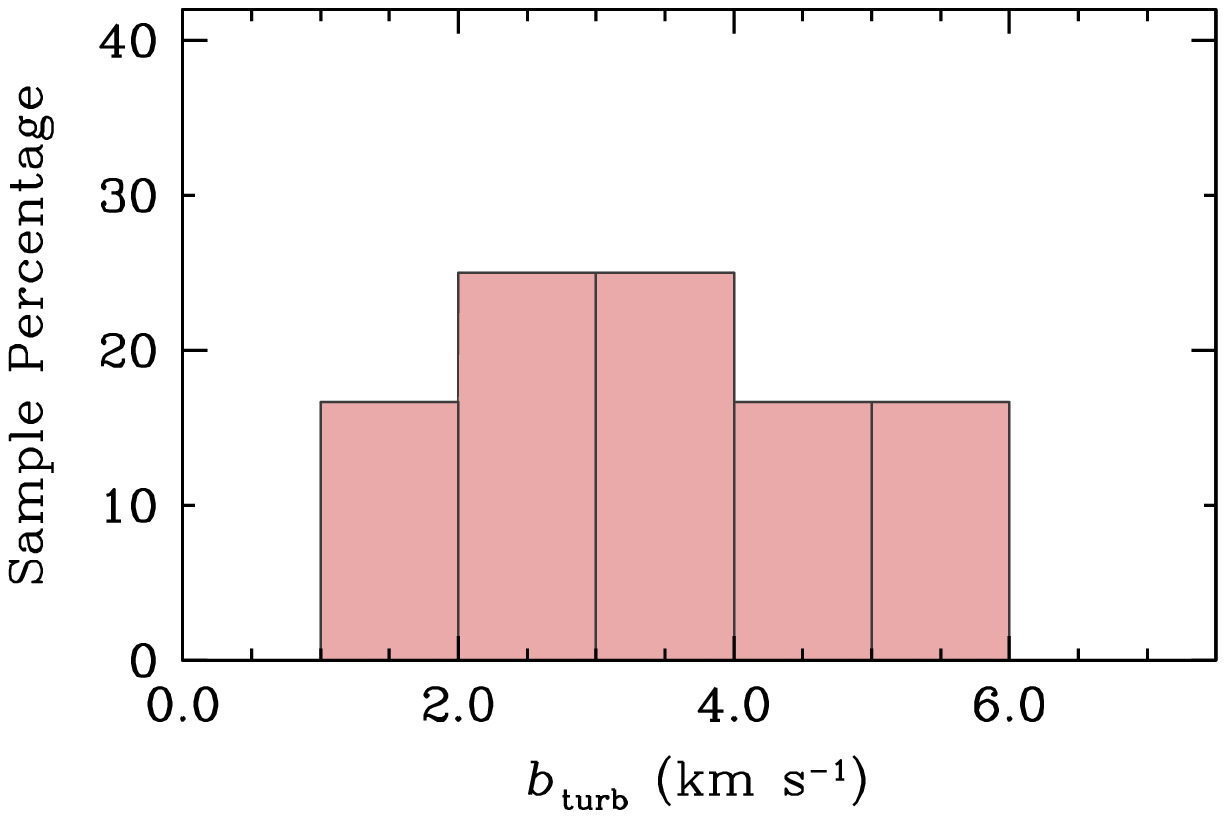}
  \hspace{0.5cm}
  \includegraphics[angle=0,width=8.5cm]{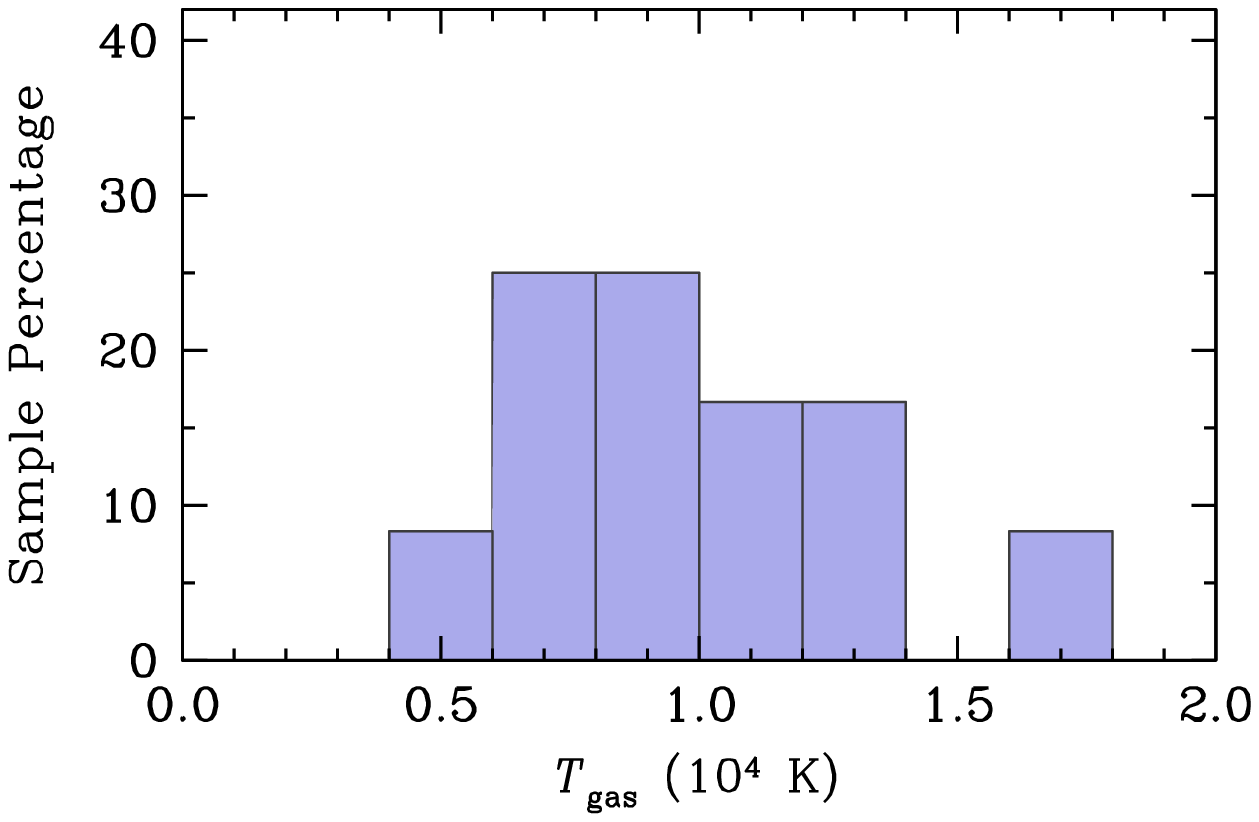}
  \caption{ 
  The distribution of the measured turbulent Doppler parameters (left panel)
  and gas kinetic temperatures (right panel) for a subset of our metal-poor DLA
  sample where the turbulent and thermal broadening of the absorption line
  profiles could be decoupled.
  }
  \label{fig:bthist}
\end{figure*}

\section{The physical properties of metal-poor DLAs}
\label{sec:properties}

As explained in the Introduction, 
the most metal-poor DLAs offer a rare opportunity to
study the physical conditions of neutral gas in the metal-poor
regime. Such conditions may provide important clues to better understand
the physics of star formation in the lowest mass dwarf galaxies at
early times. In this section, we consider some of the key physical
parameters that can be deduced from the analysis of the 
absorption lines from the metal-poor DLAs in our sample.
While there have been previous studies of this kind
\citep[e.g.][]{Sri05,WolProGaw03}, none so far have
focussed specifically on very metal-poor DLAs.

Our analysis starts with the determination of the kinetic
temperature of the gas. As explained below, this is only
possible for a subset of the metal-poor DLAs in Table~\ref{tab:kinsife}.
Specifically, only in nine out of the 23 DLAs in our sample,
can the thermal (i.e. microscopic) contribution to the line
broadening be reliably separated from the 
turbulent (i.e. macroscopic) broadening; in the other
14 cases either turbulent broadening completely dominates
or there is insufficient information in the profiles of the absorption lines
covered to decouple the two.
This inevitably introduces a bias: The physical properties 
we deduce below apply to the relatively quiescent DLAs in 
our survey, and may not be fully representative of, for example,
DLAs of a given metallicity.
Despite this bias, however, the analysis is still worthwhile,
given how little is known at present about some
of the most basic physical properties of the most metal-poor DLAs.
The sample considered in this section consists of 12 absorption
components in nine DLAs whose relevant parameters are 
collected in Table~\ref{tab:mphysprop}. Formally, the absorption
system at $z_{\rm abs}=2.536509$ towards HS\,0105$+$1619 is
a sub-DLA, exhibiting a total \HI\ column density less than the DLA
threshold of $10^{20.3}$~\HI~atoms~cm$^{-2}$. We nevertheless include this
system in our sample of DLAs, as it provides a useful indication
of the physical properties of neutral gas at somewhat lower \HI\
column density.

For several DLAs, we report the \HI\ column
density of individual components. In these rare cases, we
have detected multiple neutral deuterium lines. By assuming
that each absorption component in these systems has an identical
D/H ratio \citep{Coo14}, we have derived the \HI\ column density
of each component.

\subsection{Measured physical quantities}

DLA absorption line profiles
are characterized by a Doppler width for the line broadening, a column
density of each absorbing ion, and a redshift. The Doppler broadening further
consists of a turbulent and a thermal component, specifically:
\begin{equation}
b^2 = b_{\rm turb}^2 + \frac{2\,k_{\rm B}\,T_{\rm gas}}{m_{\rm ion}}
\end{equation}
where $k_{\rm B}$ is the Boltzmann constant, $T_{\rm gas}$ is the gas
kinetic temperature, and $m_{\rm ion}$ is the mass of the ion giving rise
to an absorption line. Thus, by considering the absorption lines from
multiple ions of widely differing mass, one can decouple the
relative contributions of turbulent and thermal broadening
(see Appendix~\ref{sec:profanal} for further details).

In columns 4 and 5 of Table~\ref{tab:mphysprop}, 
we list the derived turbulent Doppler parameters
and kinetic temperatures in the 12 absorption components 
of the nine very metal-poor DLAs
where this decoupling is possible.
For completeness, we also provide
the measured \HI\ column density in each component and the
corresponding chemical abundances. 
Our sample contains absorption systems
with a range of \HI\ column densities
($19.41\le\log N(\textrm{H\,{\sc i}})/{\rm cm}^{-2}\le20.43$)
and metallicities ($-2.64\le{\rm [O/H]}\le-1.47$).
In Figure~\ref{fig:bthist}, we show the distributions of 
turbulence parameter $b_{\rm turb}$
and of gas kinetic temperature $T_{\rm gas}$. The
weighted mean values and $1 \sigma$ dispersion
for this sample are:
$\langle b_{\rm turb} \rangle=3.5 \pm 1.1$~km~s$^{-1}$, and
$\langle T_{\rm gas} \rangle=9800 \pm 1200$~K respectively.
Despite covering an order of magnitude in both metallicity
and \HI\ column density, we found no discernible correlations
of the gas turbulence or temperature with either of these parameters.

The mean values of $b_{\rm turb}$ and $T_{\rm gas}$ reported
here for metal-poor DLAs are not dissimilar to those typical of the
so-called `Warm Neutral Medium' (WNM) component of the Milky Way
interstellar medium (ISM) today \citep{Wol95, Wol03},
although the temperatures listed in Table~\ref{tab:mphysprop}
tend to be at the upper end of the values found locally.
This broad similarity is perhaps surprising, given the 
difference by 2--3 orders of magnitude in metallicity
between the high-$z$ DLAs and the present-day 
Galactic ISM, and the attendant differences in the dust-to-gas
ratio and in the abundance of molecules \citep{PetSriLed00}.
Furthermore, the DLA gas is likely to be exposed
to a radiation field which is different in both shape and intensity
from that irradiating local interstellar clouds.
These are all factors
that regulate the heating and cooling balance of the ISM;
thus, the similarities found here between high-$z$
metal-poor DLAs and the local
WNM warrant a re-examination of two-phase models of the 
ISM under the conditions that apply at $z~=~2-3$.

Only a small number of studies have previously attempted to
measure the temperature of DLA gas
at high redshift. In a few favorable cases,
a curve-of-growth analysis has been used to
estimate the Doppler broadening of unresolved absorption
lines from molecular or neutral metal absorption lines \citep{Jor09,Tum10}.
In both of these studies, the temperature
of the gas was measured to be a few 100~K or less, consistent
with the typical values estimated for the `Cold Neutral Medium'
(CNM) of the Milky Way ISM. Other investigations have instead
used the relative strengths of the fine-structure states of C and Si
to estimate the DLA gas kinetic temperature \citep{How05,Jor10}.
These authors also report temperatures less than a few 100~K.
A conclusion that is emphasised by all of the above studies
is that the absorption from molecules, neutral species, and the
fine structure states provide a probe of the physical properties
of the CNM. In our work, we have directly measured the kinetic
temperature of the \HI\ gas associated with the first ions, and
have found that this DLA gas is much warmer than the CNM,
and is consistent with a WNM (see also, \citealt{Car12}).

\subsection{Derived physical quantities}

\begin{figure}
  \centering
  \includegraphics[angle=0,width=8.5cm]{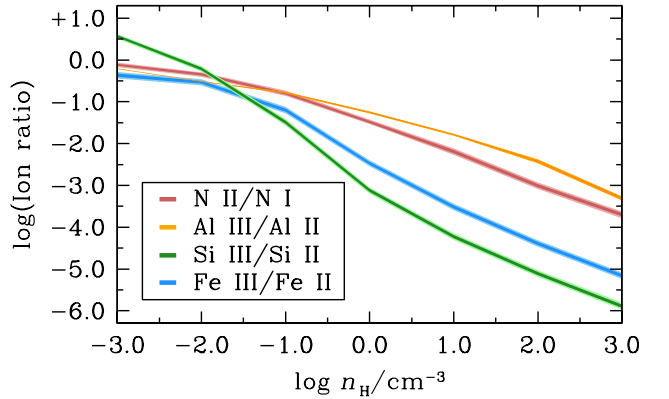}
  \caption{ 
  Solid curves show the results of our \textit{Cloudy} simulations
  of a metal-poor DLA irradiated by the \citet{HarMad12}
  UV background. This example calculation is for J0035$-$0918
  (see Appendix~\ref{sec:newsys}), which has
  $\log N({\rm Si\,\textsc{iii}})/N({\rm Si\,\textsc{ii}})=-1.08\pm0.33$, and
  $\log N({\rm N\,\textsc{ii}})/N({\rm N\,\textsc{i}})\le -0.62\,\,(3\sigma)$.
  The corresponding estimates of the H volume density based on these
  calculations are $\log n_{\rm H}/{\rm cm}^{-3} = -1.3 \pm 0.3$ and $\ge-1.4$ respectively.
  }
  \label{fig:cloudy}
\end{figure}

\begin{table*}
\centering
\begin{minipage}[c]{0.95\textwidth}
    \caption{\textsc{Derived physical properties of metal-poor DLAs$^{\rm a}$}}
    \begin{tabular}{@{}lccccccccc}
    \hline
   \multicolumn{1}{c}{\multirow{2}{*}{QSO name}}
& \multicolumn{1}{c}{\multirow{2}{*}{$z_{\rm em}$}}
& \multicolumn{1}{c}{\multirow{2}{*}{$z_{\rm abs}$}}
& \multicolumn{1}{c}{$\log n_{\rm H}$}
& \multicolumn{1}{c}{$r_{\rm H\,\textsc{i}}$}
& \multicolumn{1}{c}{$M_{\rm WNM}$}
& \multicolumn{1}{c}{$P_{\rm th}/k_{\rm B}$}
& \multicolumn{1}{c}{$P_{\rm turb}/k_{\rm B}$}
& \multicolumn{1}{c}{$c_{\rm s}$}
& \multicolumn{1}{c}{\multirow{2}{*}{${\cal M}_{\rm turb}$}}\\
    \multicolumn{1}{c}{}
& \multicolumn{1}{c}{}
& \multicolumn{1}{c}{}
& \multicolumn{1}{c}{(cm$^{-3}$)}
& \multicolumn{1}{c}{(pc)}
& \multicolumn{1}{c}{($10^{5}\,M_{\odot}$)}
& \multicolumn{1}{c}{(K\,cm$^{-3}$)}
& \multicolumn{1}{c}{(K\,cm$^{-3}$)}
& \multicolumn{1}{c}{$({\rm km\,s}^{-1}$)}
& \multicolumn{1}{c}{}\\
  \hline

\multirow{2}{*}{J0035$-$0918}  &  \multirow{2}{*}{2.420} &  \multirow{2}{*}{2.340097}  &    \multirow{2}{*}{$-1.3\pm0.3$}    &     \multirow{2}{*}{$1270^{+930}_{-610}$}  &   \multirow{2}{*}{$220^{+360}_{-160}$}  &  \multirow{2}{*}{$700^{+730}_{-360}$}  &  \multirow{2}{*}{$17^{+24}_{-10}$}   &  \multirow{2}{*}{$7.2^{+0.7}_{-0.8}$}  &  \multirow{2}{*}{$0.19^{+0.07}_{-0.07}$} \\
                               &                          &                                &                              &                             &                            &        &&&                      \\
\hline

\multirow{2}{*}{HS\,0105$+$1619}  &  \multirow{2}{*}{2.652} &  \multirow{2}{*}{2.536509}  &    \multirow{2}{*}{$-0.8\pm0.3$}   & \multirow{2}{*}{$43^{+42}_{-21}$}  & \multirow{2}{*}{$\le0.4$}   &  \multirow{2}{*}{$2790^{+2250}_{-1350}$}  &  \multirow{2}{*}{$270^{+250}_{-130}$}  &  \multirow{2}{*}{$8.51^{+0.08}_{-0.08}$}  &  \multirow{2}{*}{$0.43^{+0.02}_{-0.02}$}   \\
                               &                          &                                &                              &                             &                            &        &&&                      \\
\hline

Q0913$+$072         &  \multirow{4}{*}{2.785}  &  \multirow{2}{*}{2.618435}  &    \multirow{2}{*}{$-0.9\pm0.2$}    &   \multirow{2}{*}{$200^{+120}_{-70}$}   &  \multirow{2}{*}{$1.6^{+2.5}_{-1.0}$}  &   \multirow{2}{*}{$1440^{+890}_{-550}$}  &   \multirow{2}{*}{$380^{+210}_{-140}$}   & \multirow{2}{*}{$7.5^{+0.6}_{-0.6}$}   & \multirow{2}{*}{$0.73^{+0.07}_{-0.06}$}    \\
(comp. A)           &                          &      &    &    &    &   &   &&  \\
Q0913$+$072     &    &   \multirow{2}{*}{2.618289}  &    $-1.12\pm0.05$     &   \multirow{2}{*}{$150^{+20}_{-20}$}   &  \multirow{2}{*}{$0.37^{+0.13}_{-0.10}$}   &  \multirow{2}{*}{$640^{+80}_{-70}$}  &   \multirow{2}{*}{$140^{+20}_{-20}$}   & \multirow{2}{*}{$7.19^{+0.09}_{-0.09}$}   & \multirow{2}{*}{$0.66^{+0.04}_{-0.04}$}    \\
(comp. B)           &                           &      &    &    &    &   &   & &   \\
\hline

\multirow{2}{*}{J1001$+$0343}  &  \multirow{2}{*}{3.198}  &  \multirow{2}{*}{3.078404}  &  \multirow{2}{*}{$\ge-1.7$}    &  \multirow{2}{*}{$\le1320$}  &  \multirow{2}{*}{$\le63$}   &  \multirow{2}{*}{$\ge370$} &  \multirow{2}{*}{$\ge50$}  &  \multirow{2}{*}{$10.7^{+2.2}_{-2.8}$}  &    \multirow{2}{*}{$0.52^{+0.17}_{-0.11}$}  \\
                               &                          &                                &                              &                             &                            &        && &                     \\
\hline

\multirow{2}{*}{J1111$+$1332}  &  \multirow{2}{*}{2.420}  &  \multirow{2}{*}{2.270940}  &    \multirow{2}{*}{$-1.2\pm0.3$}    &  \multirow{2}{*}{$1080^{+890}_{-530}$}  &  \multirow{2}{*}{$180^{+340}_{-130}$}   &  \multirow{2}{*}{$1440^{+1440}_{-730}$}  &    \multirow{2}{*}{$56^{+58}_{-28}$}  &    \multirow{2}{*}{$9.3^{+0.7}_{-0.7}$}  &    \multirow{2}{*}{$0.28^{+0.04}_{-0.03}$}  \\
                               &                          &      &                              &                             &                            &              && &               \\
\hline

\multirow{2}{*}{J1358$+$6522}  &  \multirow{2}{*}{3.173}  &  \multirow{2}{*}{3.0672594}  &    \multirow{2}{*}{$-0.93\pm0.05$}   &    \multirow{2}{*}{$240^{+50}_{-40}$}  &     \multirow{2}{*}{$2.4^{+1.5}_{-0.9}$}   &     \multirow{2}{*}{$750^{+170}_{-160}$}  &    \multirow{2}{*}{$100^{+40}_{-30}$}  &    \multirow{2}{*}{$6.1^{+0.6}_{-0.6}$}  &    \multirow{2}{*}{$0.48^{+0.12}_{-0.11}$}   \\
                               &                          &     &    &    &   &  & & &  \\
\hline

J1419$+$0829$^{\rm b}$    &  \multirow{4}{*}{3.030}  &  \multirow{2}{*}{3.049654}  &    \multirow{2}{*}{$-0.35\pm0.08$}      &    \multirow{2}{*}{$32^{+7}_{-6}$}  &    \multirow{2}{*}{$0.021^{+0.010}_{-0.007}$}   &  \multirow{2}{*}{$4850^{+970}_{-820}$}  &     \multirow{2}{*}{$450^{+120}_{-90}$} &     \multirow{2}{*}{$8.07^{+0.04}_{-0.04}$}  &     \multirow{2}{*}{$0.43^{+0.03}_{-0.03}$}       \\
(comp. A)           &                          &                      &   &    &   &  &&&    \\
J1419$+$0829$^{\rm b}$  &  &  \multirow{2}{*}{3.049843}  &  \multirow{2}{*}{$\ge-1.20$}   &  \multirow{2}{*}{$\le420$}  &  \multirow{2}{*}{$\le6.2$}   &  \multirow{2}{*}{$\ge790$}  &  \multirow{2}{*}{$\ge210$}  &  \multirow{2}{*}{$8.82^{+0.08}_{-0.08}$}  &    \multirow{2}{*}{$0.74^{+0.03}_{-0.03}$}  \\
(comp. B)           &                          &                                &                              &                             &                            &                 &&&             \\
\hline

J1558$-$0031        &  \multirow{4}{*}{2.823} &  \multirow{2}{*}{2.702318}  &    \multirow{2}{*}{$-0.53\pm0.11$}   &    \multirow{2}{*}{$110^{+40}_{-30}$}   &   \multirow{2}{*}{$0.76^{+0.85}_{-0.40}$}    &  \multirow{2}{*}{$2480^{+820}_{-630}$}  &    \multirow{2}{*}{$520^{+160}_{-130}$}   &    \multirow{2}{*}{$6.9^{+0.5}_{-0.5}$}   &    \multirow{2}{*}{$0.65^{+0.06}_{-0.05}$}    \\
(comp. A)           &                          &     &    &    &                   &   &&&  \\
J1558$-$0031       & &  \multirow{2}{*}{2.702422}  &    \multirow{2}{*}{$-1.17\pm0.20$}   &    \multirow{2}{*}{$620^{+420}_{-250}$}  &    \multirow{2}{*}{$39^{+90}_{-27}$}  &     \multirow{2}{*}{$850^{+510}_{-320}$}  &    \multirow{2}{*}{$26^{+20}_{-12}$}  &    \multirow{2}{*}{$7.9^{+0.4}_{-0.4}$}  &    \multirow{2}{*}{$0.23^{+0.05}_{-0.05}$}    \\
(comp. B)           &                          &     &                              &    &                &        &&&        \\
\hline

\multirow{2}{*}{Q2206$-$199}  &  \multirow{2}{*}{2.559}  &  \multirow{2}{*}{2.076229}  &    \multirow{2}{*}{$\ge-1.6$}    &  \multirow{2}{*}{$\le1700$}  &  \multirow{2}{*}{$\le170$}   &  \multirow{2}{*}{$\ge330$} &  \multirow{2}{*}{$\ge90$}  &  \multirow{2}{*}{$9.0^{+1.1}_{-1.2}$}  &    \multirow{2}{*}{$0.72^{+0.11}_{-0.08}$}  \\
                               &                          &                                &                              &                             &                            &        && &                     \\
\hline

  \hline
    \end{tabular}

    \smallskip
    
\hspace{0.01cm}{Quoted errors are $1\sigma$ confidence intervals. Note that the probability distributions show a high degree of asymmetry.}

\smallskip

\hspace{0.01cm}$^{\rm a}${The DLAs in this table are a subset of our full sample (see Section~\ref{sec:properties}). The table columns have the following meaning:
Columns 1 and 2: Quasar name and redshift;
Column 3: Absorption redshift of the intervening DLA;
Column 4: Logarithmic volume number density of the DLA gas;
Columns 5 and 6: Radial extent and total mass of the DLA's neutral gas;
Columns 7 and 8: Thermal and turbulent pressure of the gas in units of the Boltzmann constant;
Columns 9 and 10: Sound speed and turbulent Mach number of the DLA gas.
}

\smallskip

\hspace{0.01cm}{$^{\rm b}$We caution the reader that this system is a `proximate' DLA,
with $z_{\rm abs} \simeq z_{\rm em}$. Our \textit{Cloudy} modeling does not include the radiation field from the QSO; the effect of the QSO depends on the unknown distance between it and the DLA. 
If the QSO's radiation field contributes significantly, $n_{\rm H}$ for Component A should be considered as an upper limit, whereas $n_{\rm H}$ for Component B would be unconstrained by the present models.}\\
    
    \label{tab:dphysprop}
\end{minipage}
\end{table*}

Temperature, turbulence, neutral gas column density and 
chemical composition are all parameters that can be
measured directly from the absorption lines in our spectra.
Armed with this information, we can also deduce several
other physical properties relevant to metal-poor DLAs.

To this end, we have performed a suite of
\textit{Cloudy} photoionization simulations \citep{Fer13}.
We model each DLA in our sample as a constant density
plane-parallel slab of gas irradiated by the \citet{HarMad12} extragalactic
background and the Cosmic Microwave Background at the appropriate
redshift. We consider a uniform grid of H volume densities in the range
$-3.0 \leq \log n_{\rm H}/{\rm cm}^{-3} \leq +3.0$, with a cosmic abundance
of He (i.e. $n_{\rm He}/n_{\rm H}=1/12$) and globally scale the metal
abundance of the slab to the [Si/H] abundance derived for each
DLA\footnote{Quantitatively indistinguishable results were found when
we considered a modest $\alpha$-enhancement.}.
The calculations are stopped when the \HI\ column density of the slab
is equal to that of the DLA, and a selection of metal ion column densities
are output. We perform 1000 Monte Carlo realizations of the above slab,
each time drawing a new value for the stopping \HI\ column density from
the observed value and its associated error. 
The results from an example calculation are shown in Figure~\ref{fig:cloudy}
where, for an assumed radiation field, the column density ratio of successive
ion stages of a given element is sensitive to the H volume density of the gas,
$n$(H).\footnote{We also used these simulations to confirm that ionization
corrections to the absolute metal abundances of the DLAs listed in
Table~\ref{tab:mphysprop} are $<0.1$~dex \citep[see also][]{Coo11b}.
For the sub-DLA HS\,0105$+$1619 the \textit{absolute} metallicity
correction is $-0.2$~dex, however, the ionization corrections for the
\textit{relative} metal abundance of the first ions are $<0.05$~dex.}

In metal-poor DLAs, the most commonly observed ion ratios
include \NII/\NI, \AlIII/\AlII, \SiIII/\SiII, and \FeIII/\FeII.
As can be seen from Figure~\ref{fig:cloudy},
the most sensitive probe of the H volume density is the \SiIII/\SiII\, ratio.
Which of the four ion ratios is available in a given DLA
depends on redshift, metallicity, column density, and 
blending with other absorption lines. In cases 
where more than one ion ratio is covered, the values 
of $n_{\rm H}$ deduced are generally in good mutual agreement.
Column 4 of Table~\ref{tab:dphysprop} lists the values
of DLA volume density so derived; 
$n_{\rm H} \sim 0.1$\,cm$^{-3}$
seems typical.

With these values of $n_{\rm H}$,
we can straightforwardly calculate the mass density of each DLA 
\begin{equation}
\rho_{\rm gas}=1.33\,m_{\rm p}\,n_{\rm H}
\end{equation}
where $m_{\rm p}$ is the mass of an H atom, and the factor $1.33$ accounts
for He. Using the measured kinetic temperature and turbulence of the DLAs,
together with our determinations of the number and mass densities
of the gas, the thermal and turbulent gas pressures are respectively
given by:
\begin{equation}
P_{\rm th}=(n_{\rm H}+n_{\rm He})\,k_{\rm B}\,T_{\rm gas}\simeq1.083\,n_{\rm H}\,k_{\rm B}\,T_{\rm gas}
\end{equation}
and
\begin{equation}
P_{\rm turb}=\frac{\rho_{\rm gas}\,\langle v^{2}\rangle}{2} = 
\frac{3\rho_{\rm gas}\,b_{\rm turb}^2}{4} \, .
\end{equation}
We can also
convert our measure of the kinetic temperature of the gas into
the sound speed of the gas
\begin{equation}
\label{eq:soundspeed}
c_{\rm s} = \sqrt{\frac{P_{\rm th}}{\rho_{\rm gas}}} = \sqrt{\frac{k_{\rm B}\,T_{\rm gas}}{1.23\,m_{\rm p}}}
\end{equation}
from which we can calculate the turbulent Mach number
\begin{equation}
\label{eq:mach}
{\cal M}_{\rm turb}=\frac{\sqrt{\langle{v^{2}}\rangle}}{c_s}=
\frac{\sqrt{3/2}\,b_{\rm turb}}{c_{\rm s}} \, .
\end{equation}

Note that Equations~\ref{eq:soundspeed} and \ref{eq:mach} are simply
conversions of measured physical quantities, and therefore do not
depend on the density derived from our \textit{Cloudy} simulations.

The cloud size (along the line of sight)
is obtained directly from the ratio
of the column density of \HI\ and the total H volume density:
\begin{equation}
L_{\rm H\,\textsc{i}}=\frac{N({\rm H\,\textsc{i}})}{n_{\rm H}}
\end{equation}
where we have assumed that $n_{\rm H\,\textsc i}~\simeq~n_{\rm H}$.
Our \textit{Cloudy} simulations justify this assumption, since the neutral
H volume density is $n_{\rm H\,\textsc i}~\gtrsim~0.9\,n_{\rm H}$
for gas seen in absorption as a DLA.

In order to proceed further, we need to make an assumption
about the geometry of the absorbers. One possibility
is that, by selecting for analysis the DLAs with the
lowest turbulent velocities, we preferentially pick out
face-on disks. In such cases, it is hard to extrapolate
from the parameters measured along our lines of sight 
to the global properties of the disk galaxy intersected.
On the other hand, if the gas distribution is not highly flattened,
then our data may be more representative of the 
DLA host as a whole. Thus, it is instructive to consider
the implications of our measurements for the case
of a spherical geometry of the absorber.

If the gas is spherically distributed, and our line of sight is close to the
maximum projected \HI\ column density, then the radius of these clouds
is approximately given by $r_{\rm H\,\textsc{i}}=L_{\rm H\,\textsc{i}}/2$. 
As can be seen from column 5 of Table~\ref{tab:dphysprop},
the radii of the DLAs sampled are typically less
than 1\,kpc and range from $\sim 30$ to  $\sim 1300$\,pc. 
Radius and density together give the total mass of warm
neutral gas (again, assuming a spherical geometry):
\begin{equation}
M_{\rm WNM}=\rho_{\rm gas}\times\frac{4\pi}{3}\,r_{\rm H\,\textsc{i}}^{3}
\end{equation}
which is a lower limit if the quasar
line of sight does not pass through the center of the metal-poor DLA.
All of the quantities derived above are listed in Table~\ref{tab:dphysprop}.

\begin{figure}
  \centering
  \includegraphics[angle=0,width=8.5cm]{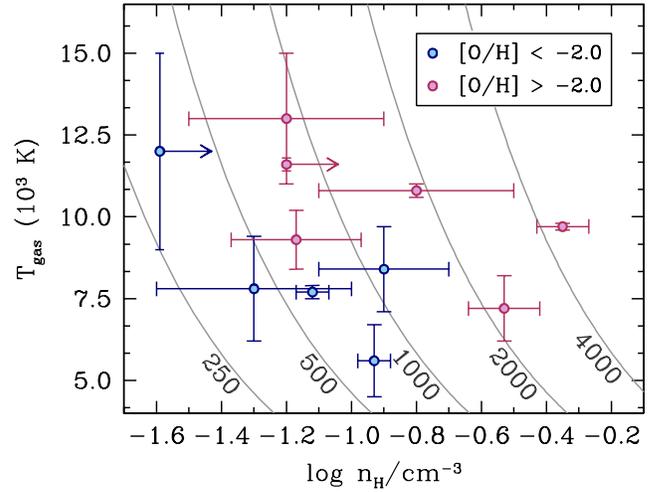}
  \caption{ 
  The relationship between the kinetic temperature $T_{\rm gas}$
  and the H volume  density $n_{\rm H}$  of metal-poor DLAs. 
  We have divided our sample equally into
  a metal-richer (with [O/H]~$\gtrsim -2.0$; red points) and a
  metal-poorer sub-sample (with [O/H]~$\lesssim-2.0$; blue points).
  From this comparison, it is apparent that the relatively
  metal-richer DLAs favor somewhat higher temperatures
  and densities compared with the metal-poorer DLAs.
  The gray solid curves indicate lines of constant
  pressure,  labelled in units of $P_{\rm th}/k_{\rm B}$.
  For comparison, the mean thermal pressure of the neutral ISM 
  near the Sun is $P_{\rm th}/k_{\rm B} \simeq 3000 $\,K~cm$^{-3}$.
  }
  \label{fig:tnhioh}
\end{figure}

\subsection{The physical properties of near-pristine neutral gas}

The physical properties that we have derived provide new insights
into the nature of metal-poor DLAs and of star formation in the lowest
metallicity environments. We first investigate the relationship between
the kinetic temperature and H volume density, illustrated in
Figure~\ref{fig:tnhioh}.
We divide our sample equally in two, and colour code the symbols
according to their [O/H] metallicity; blue symbols represent the
relatively metal-poorer sub-sample (with [O/H]~$\lesssim-2.0$), whilst the
red symbols are for the relatively metal-richer sub-sample
(with [O/H]~$\gtrsim-2.0$). In Figure~\ref{fig:tnhioh}, we have also drawn
a series of solid lines of constant pressure
($P_{\rm th}/k_{\rm B}$)
in the $T_{\rm gas}$--$n_{\rm H}$ plane. 
Although the sample size is still 
small, these data suggest that metal-richer DLAs have a slight
preference to host gas that is at modestly higher densities
and temperatures, and therefore higher pressures, compared
with the metal-poorer sample.

We note that the majority of the DLAs in our sample are at lower pressures
than the mean $P_{\rm th}/k_{\rm B} \simeq 3000$\,K~cm$^{-3}$
determined by \citet{JenTri01,JenTri11} for diffuse 
interstellar clouds within a few kpc
of the Sun.  Pressures in metal-poor DLAs are also significantly lower 
than those typical of  DLAs with detectable 
column densities of molecular hydrogen, which tend to be 
at the metal-rich end of the distribution of DLA
metallicities \citep{Sri05}.

\begin{table}
\centering
\begin{minipage}[c]{0.38\textwidth}
    \caption{\textsc{Summary of the physical properties of metal-poor DLAs}}
    \begin{tabular}{@{}lc}
    \hline
   \multicolumn{1}{l}{Property}
& \multicolumn{1}{c}{Confidence Intervals} \\
  \hline

\hline

\multirow{2}{*}{$b_{\rm turb}$ (km s$^{-1})$}   & \multirow{2}{*}{$3.3^{+1.8}_{-1.5}$ ($1\sigma$) $^{+2.3}_{-2.4}$ ($2\sigma$)}\\
&\\
\multirow{2}{*}{$T_{\rm gas}$ (K)}   & \multirow{2}{*}{$9600^{+2500}_{-2600}$ ($1\sigma$) $^{+12200}_{-5000}$ ($2\sigma$)}\\
&\\
\multirow{2}{*}{$\log n(\textrm{H})/{\rm cm}^{-3}$}   & \multirow{2}{*}{$-1.0^{+0.4}_{-0.3}$ ($1\sigma$) $^{+0.6}_{-0.7}$ ($2\sigma$)}\\
&\\
\multirow{2}{*}{$r_{\rm H\,\textsc{i}}$  (pc)} & \multirow{2}{*}{$220^{+840}_{-130}$ ($1\sigma$) $^{+2200}_{-200}$ ($2\sigma$)}\\
&\\
\multirow{2}{*}{$\log M_{\rm WNM}/{\rm M}_{\odot}$} & \multirow{2}{*}{$5.4^{+1.9}_{-0.9}$ ($1\sigma$) $^{+2.5}_{-2.5}$ ($2\sigma$)}\\
&\\
\multirow{2}{*}{$P_{\rm th}/k_{\rm B}$  (K cm$^{-3}$)} & \multirow{2}{*}{$1050^{+1610}_{-450}$ ($1\sigma$) $^{+4150}_{-730}$ ($2\sigma$)}\\
&\\
\multirow{2}{*}{$P_{\rm turb}/k_{\rm B}$  (K cm$^{-3}$)} & \multirow{2}{*}{$130^{+330}_{-100}$ ($1\sigma$) $^{+630}_{-120}$ ($2\sigma$)}\\
&\\
\multirow{2}{*}{$c_{\rm s}$  (km s$^{-1}$)} & \multirow{2}{*}{$8.0^{+1.0}_{-1.2}$ ($1\sigma$) $^{+4.1}_{-2.5}$ ($2\sigma$)}\\
&\\
\multirow{2}{*}{${\cal M}_{\rm turb}$} & \multirow{2}{*}{$0.50^{+0.23}_{-0.24}$ ($1\sigma$) $^{+0.34}_{-0.36}$ ($2\sigma$)}\\
&\\

  \hline
    \end{tabular}
    
    \label{tab:sumpp}
\end{minipage}
\end{table}


By inspection of the thermal and turbulent pressures listed in
Table~\ref{tab:dphysprop}, we can also conclude that the most
metal-poor DLAs are predominantly held up by thermal pressure,
as is the case for the WNM in the local ISM \citep{Val96,RedLin04}.
Similarly, the turbulent Mach number that we derive for the DLAs
in our sample (${\cal M}_{\rm turb}\sim0.5$) is reminiscent of the local ISM values;
every DLA that we have analyzed here exhibits a subsonic
turbulent Mach number. Such low Mach numbers suggest that the
WNM in the most metal-poor DLAs follows a fairly smooth gas density
distribution \citep{KowLazBer07,McKOst07}, which is not conducive
to gas fragmentation. Presumably, star formation in the most metal-poor
DLAs proceeds via a cold neutral medium which, based on local studies
\citep[e.g.][]{HeiTro03}, would exhibit a supersonic ${\cal M}_{\rm turb}$.
The existence of a two-phase medium with a cold and warm
component is supported by \HI\ 21\,cm observations of neutral gas in
Leo T and Leo P --- two nearby low metallicity dwarf galaxies in the
Local Group \citep{Rya08,Ber14}.

Such observations have not yet been conducted on redshift $z\sim3$ very metal-poor
DLAs. However, there exists a handful of 21\,cm absorption measurements
for DLAs in front of radio-bright quasars \citep{Sri12,Kan14}. In this case, the optical
depth of 21\,cm absorption depends on the \HI\ column density, the spin temperature,
and a covering factor that describes how effectively the foreground DLA covers the
extended radio emission of the background quasar. For gas in a WNM, \citet{Lis01}
demonstrated that collisional excitation
is unable to efficiently thermalize the \HI\ $\lambda$21\,cm line, leading to spin
temperatures that are somewhat less than the kinetic temperature of the gas.
A relationship between the spin and kinetic temperatures is further
complicated by the uncertain fraction of gas along the line-of-sight that is in a
CNM versus a WNM; a high CNM fraction weights the line-of-sight spin temperature
towards lower values. In principle, the fewer coolants that are available in low
metallicity gas imply lower CNM fractions, and therefore higher spin temperatures;
presumably this is the origin of the observed anti-correlation between spin temperature
and metallicity \citep{KanChe01,Kan14}. By extrapolating the relationship between
spin temperature and metallicity of the DLA population, the most metal-poor DLAs in
our analysis presumably exhibit relatively high spin temperatures, corresponding to
a low fraction of CNM gas.

For convenience, we provide a summary of the key physical
properties of the most metal-poor DLAs in Table~\ref{tab:sumpp}\footnote{For
the physical properties that depend on the gas volume density, we found that
the values changed by $\sim10$ per cent when we exclude the proximate DLA
towards J1419$+$0829. In addition, we have found that the sub-DLA towards
HS\,0105$+$1619 has very similar properties to the DLA sample; the inclusion
or exclusion of this sub-DLA has a negligible affect on the average properties
listed in Table~\ref{tab:sumpp}.}.
In brief, our observations and analysis suggest that the
most metal-poor DLAs are clouds that typically contain a
reservoir of $\sim10^{4-7}\,{\rm M}_{\odot}$ of warm neutral gas
with a temperature $T_{\rm gas}\simeq9600$\,K. This gas is
expected to have a smooth density distribution, and is predominantly
supported by thermal pressure. Although our measurement technique
is not sensitive to the presence or absence of a CNM (see
Appendix~\ref{sec:profanal} for details), we point out that the
properties of the inferred WNM are not conducive to star formation.
Given the absence of molecular absorption lines in our DLA sample,
considered a tracer of the CNM, we suggest that a CNM could
be present in these DLAs with either a small
cross-sectional area, or may only exist for a relatively short
phase of a DLA's life just prior to a burst of star formation.
The absence of a CNM in our DLA sample is consistent with
the paucity of strong molecular hydrogen absorption associated
with the general DLA population \citep{Jor14}.
In addition, metal-poor DLAs might contain very low CNM
fractions, in agreement with recent cosmological hydrodynamic
simulations \citep{MaiTesCoo14}.

\section{Metal-poor DLAs and modern-day Dwarf Galaxies}
\label{sec:disc}

In the previous sections, we have assessed the kinematic,
chemical and physical aspects of the metal-poor DLA population. 
With these observations, we can start building a connection between
the metal-poor DLA population at high redshifts
and their modern-day counterparts
which are presumably akin to Local Group dwarf galaxies \citep{SalFer12}.
In this section, we summarize what can be learnt by studying the lowest
mass galaxies over cosmic time.

Detailed star formation histories that are extracted
from color-magnitude diagrams of Local Group dwarf galaxies
\citep{Wei11,Wei14a} imply that these galaxies  formed a
considerable fraction of their present day stellar population
between redshifts $z\sim2-5$ \citep{Wei14a}, corresponding
to the range of redshifts that are probed by our metal-poor
DLA sample. Furthermore, the observed kinematics of the
metal-poor DLA population show a similar trend to that
observed in Local Group galaxies, where the systems
with the lowest metallicities also exhibit the lowest velocity
dispersions.
From this observation, we propose that the most
metal-poor DLAs are associated with the lowest
mass halos that can potentially form stars at
redshift $z\sim3$.

This conclusion receives additional support 
from the observed chemical evolution of the DLA population. 
Specifically, the evolution of the
$\alpha$/Fe ratio with metallicity (as measured by Fe/H)
for a typical DLA is broadly similar to
that seen in the local, low mass dwarf galaxies, and 
is clearly different from the behaviour of this ratio in
MW stars of the halo and thin and thick disks \citep{Ven04,Ben05}.
The chemical evolution of a typical DLA also differs, although to
a lesser extent, in the dependence of the $\alpha$/Fe ratio on metallicity compared
to relatively massive dwarf galaxies, such as the  Small and Large Magellanic Clouds
\citep{HilBarSpi97,Ven99,Pom08} and the Sagittarius dSph \citep{deB14}.
The observations presented herein suggest that the chemical evolution of a typical DLA is
similar to galaxies that are somewhat less massive than the Magellanic Clouds.
By extension, we propose that the most metal-poor DLAs at redshift $z \sim 3$
most likely represent the earliest stages of chemical evolution that were
experienced by the lowest mass dwarf galaxies.

Of course, the observed $z \sim 3$ population of metal-poor DLAs
need not represent the entire Local Group dwarf galaxy population.
For example, some of the ultra-faint dwarfs appear to have
formed most of their stars prior to $z \sim 3$
(e.g. Hercules and Leo IV; \citealt{Wei14a,Wei14b}).
Indeed, some of the UFDs may have experienced
`reionization quenching' \citep[e.g.][]{BulKraWei00}, and may
therefore not be observable as gas-rich metal-poor DLAs at
$z \sim 3$. Nevertheless, these near pristine DLAs can
still improve our understanding of the physical conditions for
star formation in this low metallicity regime.

Finally, consider the current working definition of a `first galaxy' ---  that
is, in its simplest terms, a long-lived stellar population confined by a
dark matter halo (see e.g. the review by \citealt{BroYos11}). Using this
definition, we suggest that the \textit{most} metal-poor DLAs may trace
a reservoir of largely primordial neutral gas that is confined to a dark
matter halo yet to form its first long-lived stellar population. Indeed,
some near-pristine DLAs may  contain the hallmark chemical
signatures of enrichment by metal-free Population~III stars
\citep[][the chemical abundance pattern of that DLA is updated herein -- see Appendix~\ref{sec:newsys}]{Coo11a}.
It is therefore certainly plausible that some of the most
metal-poor DLAs may not even fit the canonical definition
of a `galaxy', until they form their second generation of stars.

\section{Summary and Conclusions}
\label{sec:conc}

In this paper, we have provided the first observational links between
the local, well-studied population of dwarf galaxies and their
most likely progenitors at high redshift. Our study considers the
kinematic, chemical, and physical properties of the metal-poor
DLA population, which all favour the interpretation that such
systems predominantly trace the early evolution of dwarf
galaxies. From the sample of metal-poor DLAs considered
here, we draw the following conclusions.

\smallskip

\noindent ~~(i) We measure the kinematics of a sample
of 23 metal-poor DLAs with [Fe/H]~$\lesssim-2.0$, and
conclude that the most metal-poor systems are likely to be associated
with the lowest mass galaxies that have the potential to form stars
at redshift $z \sim 3$. We also find a close correspondence
between our DLAs and the kinematics of Local Group dwarf
galaxies of similar metallicities.

\smallskip

\noindent ~~(ii) To better understand the chemical evolution of
the whole DLA population, we have considered the variation in
[$\alpha$/Fe] of DLAs as a function of [Fe/H] metallicity. We find
evidence for a `knee'  in the evolution of [$\alpha$/Fe]
in DLAs at [Fe/H]~$\simeq-2.0$, which is entirely consistent with
the knee that is observed in most of the dwarf spheroidal
galaxies near the Milky Way (assuming [Zn/Fe]~$\simeq0$ when
[Fe/H]~$\simeq-2.0$, as supported by observations of dSphs). 
Conversely, the behaviour  of [$\alpha$/Fe] vs. [Fe/H]
in the DLA population clearly differs from that exhibited
by Milky Way halo stars. We conclude that the
$z\sim 3$ DLA population is not representative of 
galaxies similar to those that merged to form the 
bulk of the Milky Way stellar halo.

\smallskip

\noindent ~~(iii) We have examined the physical conditions of the
gas giving rise to metal-poor DLAs, by compiling a sample
of systems with the most well-determined cloud models where
it is possible to decouple the turbulent and thermal broadening
of the line profiles. From the analysis of nine metal-poor DLAs
([Fe/H]~$\sim-2.3$) with a total of 12 cloud components, we
estimate that the typical values of the turbulent Doppler parameter
and of the gas kinetic temperature are
$b_{\rm turb}=3.3^{+1.8}_{-1.5}$~km~s$^{-1}$, and
$T_{\rm gas}=9600^{+2500}_{-2600}$~K respectively.

\smallskip

\noindent ~~(iv) Using \textit{Cloudy} photoionization calculations,
we have also estimated the gas density 
in each of the 12 DLA absorbers.
Taken together, these measures have allowed us to
estimate the mass, radial extent, thermal and turbulent pressure,
and turbulent Mach number for these near-pristine clouds of
neutral gas. We find that metal-poor DLAs contain roughly
$10^{4-7}\,{\rm M}_{\odot}$ of neutral gas, and are
predominantly supported by thermal pressure. Furthermore,
all DLAs in our sample exhibit subsonic Mach numbers,
which implies that the distribution of warm neutral
gas in these systems is relatively smooth.

\smallskip

Our analysis has provided new insights into the nature of
the most metal-poor DLAs, and the first glimpse
into the physical conditions of star formation in low metallicity
gas at high redshift. This now opens up the exciting prospect
of studying in depth with future work the early evolution of 
galaxies analogous to the Local
Group dwarfs, more than 10 billion years in the past,
at a time when they were building the bulk of their present-day
stellar populations.

\section*{Acknowledgments}
We are grateful to the staff astronomers at the ESO VLT and Keck  
Observatory for their assistance with the observations.
We thank the Hawaiian people for the opportunity to observe from Mauna Kea;
without their hospitality, this work would not have been possible.
We are thankful to an anonymous referee for a constructive and
thorough report that improved the presentation of our work.
We also wish to thank Tom Abel, Thomas de Boer, Mark Krumholz,
Marc Rafelski, and Erik Tollerud for valuable discussions and comments.
M.~P.~ would like to thank Matteo Monelli and Stefania Salvadori
for organizing the Symposium ``Local Group, local cosmology''
(EWASS 2013) that provided useful insights into some of the
properties of Local Group dwarf galaxies.
R.~J.~C. was partially supported by NSF grant AST-1109447
during this work, and is currently supported by NASA through
Hubble Fellowship grant HST-HF-51338.001-A, awarded by the
Space Telescope Science Institute, which is operated by the
Association of Universities for Research in Astronomy, Inc.,
for NASA, under contract NAS5- 26555.
R.~A.~J. gratefully acknowledges support from the NSF
Astronomy and Astrophysics Postdoctoral Fellowship
under award AST-1102683.

\appendix

\section{A. Thermal and turbulent broadening}
\label{sec:profanal}

The cloud models for all of the DLAs considered in this work were analyzed
with the new Voigt profile fitting software recently developed for precision
absorption line profile analysis \citep{Coo14}. This Absorption LIne Software
(\textsc{alis}), uses a Levenberg-Marquardt chi-squared minimization algorithm to
minimize the residuals between the data and the model, weighted by the error
spectrum (e.g. \citealt{Mar09}). \textsc{alis} uses the atomic data compiled by
\citet{Mor03}, with updates from both \citet{JenTri06} and \citet{MurBer14}.

The absorption line profiles for the most metal-poor DLAs typically have
a very low velocity dispersion, often with just a single narrow absorption
line. In the present work, we have utilized the simplicity of the absorption
line profiles exhibited by the most metal-poor DLAs to decouple the
relative contributions of turbulent and thermal broadening; thermal
broadening depends inversely on the square-root of the atom's mass,
whereas turbulent broadening is the same for all gas constituents.
Thus, one can measure the kinetic temperature of the absorbing gas
cloud \textit{directly}, provided that:
(1) The gas in each absorption component is distributed according to
a Maxwell-Boltzmann distribution;
(2) an array of transitions of differing oscillator strengths from a given
ion is available to populate the curve-of-growth and thereby measure
the total broadening, and
(3) absorption lines from at least two atoms of widely differing mass
are accessible to measure the thermal contribution to the line
broadening. The ideal ions that satisfy these criteria are the
light \DI\ with its host of Lyman series transitions in combination with
one of the heavier atoms (such as \OI, \SiII, or \FeII) which have a range of
transitions with a variety of strengths and are commonly observed
in DLAs. In some favorable cases (see Appendix~\ref{sec:newsys}),
it is also possible to solely use a selection of metal absorption lines to
obtain a handle on the gas kinetic temperature \citep{Car12}.

In many cases, the combined thermal and turbulent broadening of the
metal line profiles is only marginally resolved from the instrumental
broadening. An accurate measure of the instrumental profile is therefore required in order
to pin down the cloud model of the absorbing gas. Moreover, if the point
spread function of the quasar does not uniformly illuminate the spectrograph's
slit, the instrumental resolution during the observations will be less than the
nominal instrumental resolution measured from the widths of the ThAr
arc lines. We overcome this uncertainty by assuming that the instrumental
broadening function is closely approximated by a Gaussian, and allow the
FWHM of the Gaussian profile to be a free parameter in our modeling
procedure. Therefore, our uncertainty in the instrumental profile is folded
into the uncertainty in both the cloud model and chemical abundances.
Our profile analysis also folds in the uncertainty in the placement of the
quasar continuum level, by fitting a low order polynomial to the quasar
continuum during the absorption line fitting procedure.

DLAs are generally considered to host a cold and a warm neutral medium
in pressure equilibrium \citep[e.g.][]{Wol95}. We have performed a suite of
Monte Carlo simulations to test the sensitivity of the above technique to the
presence of a two-phase medium. We consider two test cases:
(1) A system with only metal-absorption lines (J0035$-$0918), and
(2) a system with metal, \DI, and \HI\ absorption lines (J1419$+$0829).

Using the best-fitting absorption line profiles, we generated a new model
where 50 per cent of the gas along the line-of-sight resides in a CNM (100\,K),
and the remaining gas is in a WNM with the measured kinetic temperature.
100 Monte Carlo realizations were generated by perturbing these model
profiles by the error spectrum of the data. We first perform a fit to each realization
that includes both a CNM \textit{and} a WNM. On the basis of these simulations we
conclude that we are unable to recover the properties of the CNM; our technique to
measure the kinetic
temperature is therefore insensitive to the presence of a CNM. We then performed
a second fit to each realization but this time we only included a WNM component in
the fit.
With this test, we were able to recover the properties of the WNM, without any significant
bias. Therefore, if the DLAs that we probe in our sample contain a CNM, the physical
properties that we derive are largely unaffected, although we are unable to detect
the presence of a CNM. In future studies, it may be possible to identify a CNM in
metal-poor DLAs by detecting absorption from molecular hydrogen or \CI. Absorption
lines from these species are not present in our sample.

\section{B. New systems}
\label{sec:newsys}

The majority of our sample is described in our previous work
\citep{Pet08,Coo11b,Coo14}. In this Appendix, we present
our analysis of a newly discovered metal-poor DLA at
redshift $z_{\rm abs}\simeq2.271$ towards J1111$+$1332,
and provide an improved reanalysis of the metal-poor DLA
reported by \citet{Coo11a} at $z_{\rm abs}\simeq2.340$ towards
J0035$-$0918. In both of these systems, the data cover
absorption lines with a range of oscillator strengths for several
elements between C and Ni. The difference in atomic mass for
these elements, combined with the broad range of oscillator
strengths for many transitions, has allowed us to decouple the thermal
broadening of the line profile (which depends on the atomic mass)
from that of turbulent broadening (assumed to be the same for
all elements), using metal absorption lines alone.

\subsection{J0035$-$0918, $z_{\rm abs}=2.340097$}

The DLA at redshift $z_{\rm abs}\simeq2.340$ towards
the quasar J0035$-$0918 is currently one of the most
metal-poor systems known. It was initially reported by
\citet{Coo11a} as the first known example of a DLA
with an abundance pattern resembling the nucleosynthesis
expected by the first stars. In particular, this DLA was
reported to have an abnormal excess of C and N relative
to Fe, [C/Fe]=+1.53. This result was later contested by \citet{Car12},
who correctly pointed out that a contribution from thermal broadening
would result in a lower overabundance of C/Fe. Using the
\citet{Coo11a} data, these authors estimated the carbon overabundance
for this system to be  [C/Fe] $+0.51\pm0.10$ for a cloud model that is
purely broadened by thermal motions; this value is a theoretical lower limit
on the true [C/Fe] abundance for this system, which is likely broadened
by both mechanisms. Unfortunately, the initial data reported by
\citet{Coo11a} did not cover several important \NI\ and \FeII\ absorption
lines with a high oscillator strength which would have permitted
a decoupling of the thermal and turbulent motions.

\begin{table}
\centering
\caption{\textsc{Ion column densities of the DLA in J0035$-$0918 at $z_{\rm abs}=2.340097$}}
    \begin{tabular}{@{}lp{1.8in}ccc}
    \hline
  \multicolumn{1}{l}{Ion}
& \multicolumn{1}{c}{Transitions used}
& \multicolumn{1}{c}{log $N$(X)/${\rm cm}^{-2}$}
& \multicolumn{1}{c}{[X/H]}
& \multicolumn{1}{c}{[X/Fe]}\\
    \hline
\HI    &  1215  &  20.43 $\pm$ 0.04 & $\ldots$ & $\ldots$ \\
\CII   &  1036, 1334  &  14.50 $\pm$ 0.15 & $-2.36\pm0.16$ & $+0.58\pm0.16$ \\
\NI    &  1134.1, 1134.4, 1134.9, 1199.5, 1200.2, 1200.7      &  13.37 $\pm$ 0.05  & $-2.89\pm0.06$ & $+0.05\pm0.07$ \\
\NII    &  1084      &  $\le12.60^{\rm a}$ & $\ldots$ & $\ldots$ \\
\OI    &  971, 988, 1039, 1302        &  14.68 $\pm$ 0.06 & $-2.44\pm0.07$ & $+0.50\pm0.08$ \\
\AlII  &  1670  &  11.68 $\pm$ 0.06 & $-3.19\pm0.07$ & $-0.25\pm0.08$ \\
\SiII  &  989, 1193, 1260, 1304, 1526  &  13.37 $\pm$ 0.05 & $-2.57\pm0.06$ & $+0.37\pm0.07$ \\
\SiIII  &  1206  &  12.29 $\pm$ 0.33 & $\ldots$ & $\ldots$ \\
\SII  &  1259  &  13.08 $\pm$ 0.10 & $-2.49\pm0.11$ & $+0.45\pm0.11$ \\
\FeII  &  1608, 2344, 2374, 2382, 2586, 2600        &  12.96 $\pm$ 0.05  & $-2.94\pm0.06$ & $\ldots$ \\
\FeIII  &  1122        &  $\le13.02^{\rm a}$  & $\ldots$ & $\ldots$ \\
    \hline   
    \end{tabular}

{$^{\rm a}\,3\sigma$ upper limit.~~~~~~~~~~
~~~~~~~~~~~~~~~~~
~~~~~~~~~~~~~~~~~
~~~~~~~~~~~~~~~~~
~~~~~~~~~~~~~~~~~
~~~~~~~~~~~~~~~~~
~~~~~~~~~~~~~~~~~}\\
\label{tab:J0035m0918_cd}
\end{table}

By combining the \citet{Coo11a} data with the recent observations
described by \citet{Dut14}, we find that a single component at redshift
$z_{\rm abs}=2.340097\pm0.000001$ provides an excellent fit to the data,
with a turbulent Doppler parameter $b_{\rm turb}=1.1\pm0.4$ and a
temperature $T_{\rm gas}=7800\pm1600$. Using the new, high-resolution
UVES spectrum we derive an \HI\ column density from the \Lya\ transition
$\log N({\rm H\,\textsc{i}})/{\rm cm}^{-2}=20.43\pm0.04$. This estimate is
lower by $\sim0.1$~dex from that reported by \citet{Coo11a}, who used a medium resolution
spectrum from the Magellan Echellette spectrograph to derive $N(\textrm{H\,{\sc i}})$.
The overabundance of C relative to Fe for our best-fit model is [C/Fe] $=+0.58\pm0.16$,
implying that thermal broadening does indeed contribute significantly to
the cloud kinematics of this DLA. 
For compariosn, \citet{Dut14} deduced  [C/Fe]~$=+0.45\pm0.19$,
which is consistent with, but somewhat lower than, the minimum 
value estimated by \citet{Car12} for a purely thermally broadened
cloud model. Nevertheless, our analysis is consistent with their
estimate. The derived column densities for this system are collected
in Table~\ref{tab:J0035m0918_cd}. 
Overall, this DLA still exhibits an
enhancement in C/Fe, greater than that exhibited by a typical very
metal-poor DLA by $2.4\sigma$, however it no longer
meets the generally adopted criterion defining 
Carbon Enhanced Metal Poor (CEMP) stars, [C/Fe]~$\ge+0.7$
\citep{Aok07}.
Although this system is not quite as
abundant in carbon as once thought, it is certainly a chemically
peculiar system, and highlights the need to push these studies to
even lower metallicities where such anomalous abundance patterns
are expected to be more common \citep{CooMad14}.

\subsection{J1111$+$1332, $z_{\rm abs}=2.270940$}
\label{sec:J1112}

The second system reported here is the $z_{\rm abs}\simeq2.271$ DLA
towards the $m_{\rm r}\simeq17.1$ quasar J1111$+$1332. We first
recognized this DLA as an excellent candidate VMP DLA from the
relatively high S/N SDSS discovery spectrum, which exhibits strong
\HI\ \Lya\ absorption line together with the apparent absence of
associated metal absorption lines.
We performed follow-up observations of J1111$+$1332 with HIRES (\citealt{Vog94})
on 2011 March 24 and 25 (Program ID: A152Hb). We employed the $1.148$
arcsec wide C5 decker (with a nominal resolution of
$v_{\rm FWHM}~=~8.3~{\rm km\,\,s}^{-1}$) and the red cross-disperser.
We recorded $6\times2700\,{\rm s}$ exposures, covering a wavelength
range $4645-8155$ with small gaps near 5100\,\AA\ and 6700\,\AA.
We used $2\times2$ on-chip binning. The data were reduced with
\textsc{makee}\footnote{\textsc{makee} is maintained by T.~Barlow,
and is available from\\
\url{http://www.astro.caltech.edu/$\sim$tb/makee/}},
following the standard reduction steps of bias subtraction and flat fielding.
The orders were defined and traced using a quartz lamp exposure with a
pinhole decker (D5). After the orders were extracted, the data were converted
to a vacuum heliocentric wavelength frame, with reference to an exposure of
a ThAr hollow-cathode lamp. The final spectrum was combined
with \textsc{UVES\_popler}\footnote{\textsc{UVES\_popler} was written and
is maintained by M.~T.~Murphy,\\and is available from\\
\url{http://astronomy.swin.edu.au/$\sim$mmurphy/UVES\_popler/}}.
The combined signal-to-noise ratio (S/N) of the data near 6000\,\AA\ is $\sim60$.

\begin{table}
\centering
\caption{\textsc{Ion column densities of the DLA in J1111$+$1332 at $z_{\rm abs}=2.270940$}}
    \begin{tabular}{@{}lp{1.8in}ccc}
    \hline
  \multicolumn{1}{l}{Ion}
& \multicolumn{1}{c}{Transitions used}
& \multicolumn{1}{c}{log $N$(X)/${\rm cm}^{-2}$}
& \multicolumn{1}{c}{[X/H]}
& \multicolumn{1}{c}{[X/Fe]}\\
    \hline
\HI    &  1215  &  20.39 $\pm$ 0.04 & $\ldots$ & $\ldots$ \\
\CII   &  1334  &  14.72 $\pm$ 0.09 & $-2.10\pm0.10$ & $+0.17\pm0.09$ \\
\NI    &  1134.4, 1134.9, 1199.5, 1200.2, 1200.7      &  13.53 $\pm$ 0.02 & $-2.69\pm0.05$ & $-0.42\pm0.02$ \\
\OI    &  1302        &  15.16 $\pm$ 0.07 & $-1.92\pm0.08$ & $+0.35\pm0.07$ \\
\AlII  &  1670  &  12.37 $\pm$ 0.02 & $-2.46\pm0.05$ & $-0.19\pm0.02$ \\
\AlIII  &  1854, 1862  &  11.90 $\pm$ 0.03 & $\ldots$ & $\ldots$ \\
\SiII  &  1190, 1193, 1260, 1304, 1526, 1808  &  13.95 $\pm$ 0.02 & $-1.95\pm0.05$ & $+0.32\pm0.02$ \\
\SiIII  &  1206  &  13.36 $\pm$ 0.05 & $\ldots$ & $\ldots$ \\
\SiIV  &  1393, 1402  &  13.09 $\pm$ 0.02 & $\ldots$ & $\ldots$ \\
\SII  &  1259  &  13.69 $\pm$ 0.09 & $-1.84\pm0.10$ & $+0.43\pm0.09$ \\
\CrII  &  2062, 2066  &  12.00 $\pm$ 0.09 & $-2.03\pm0.10$ & $+0.24\pm0.09$ \\
\FeII  &  1121, 1125, 1143, 1144, 1608, 1611, 2260, 2374, 2382     &  13.59 $\pm$ 0.01 & $-2.27\pm0.04$ & $\ldots$ \\
\NiII  &  1317, 1370, 1454, 1751  &  12.15 $\pm$ 0.06 & $-2.45\pm0.07$ & $-0.18\pm0.06$ \\
\ZnII  &  2026  &  $\le11.08^{\rm a}$ & $\le-1.94$ & $\le+0.33$ \\
    \hline   
    \end{tabular}

{$^{\rm a}\,3\sigma$ upper limit.~~~~~~~~~~
~~~~~~~~~~~~~~~~~
~~~~~~~~~~~~~~~~~
~~~~~~~~~~~~~~~~~
~~~~~~~~~~~~~~~~~
~~~~~~~~~~~~~~~~~
~~~~~~~~~~~~~~~~~}\\
\label{tab:J1111p1332_cd}
\end{table}  

\begin{figure*}
  \centering
  \includegraphics[angle=0,width=17.0cm]{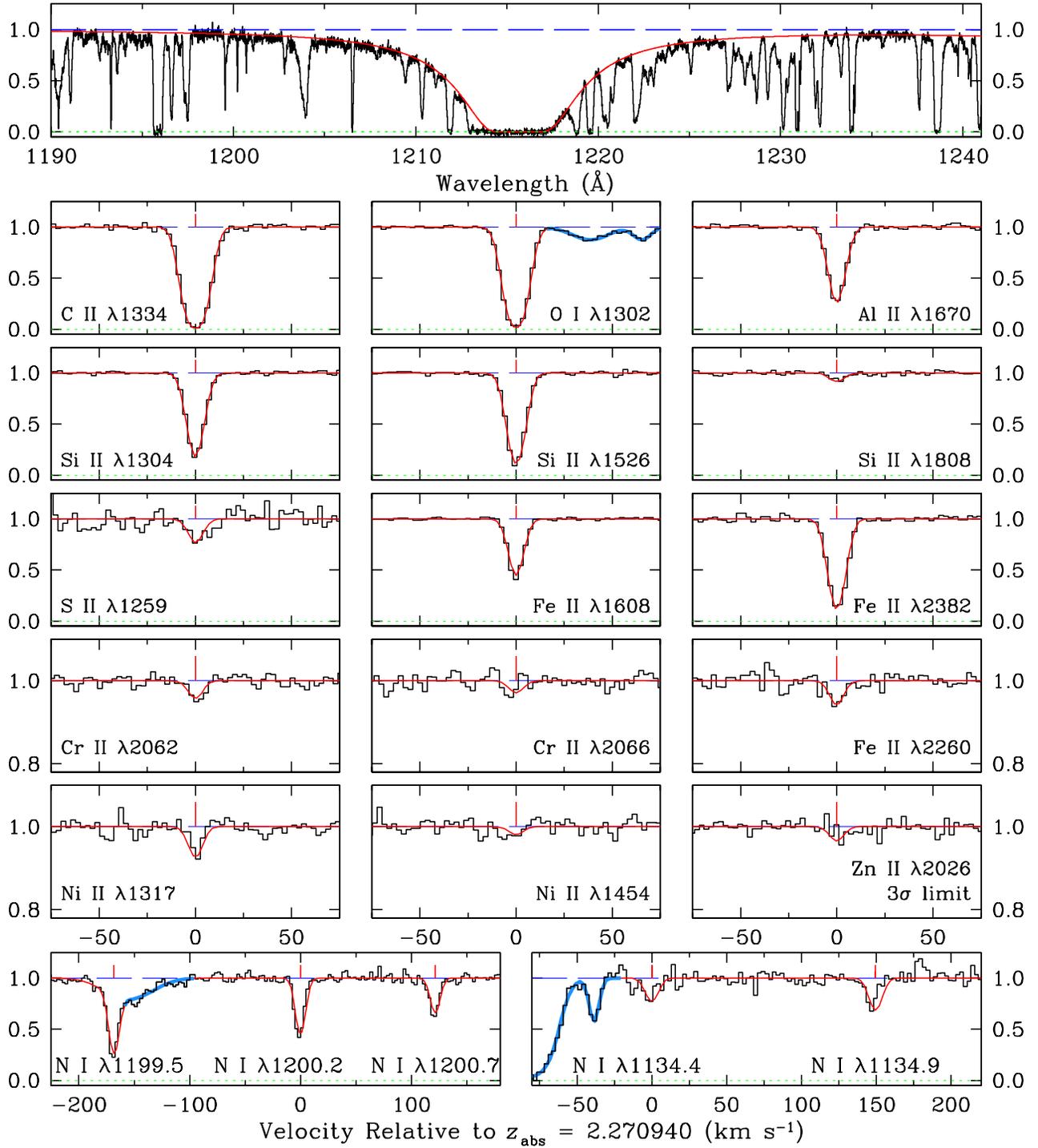}
  \caption{ 
The top panel shows the damped \Lya\ line for the DLA at $z_{\rm abs}=2.270940$
towards the QSO J1111$+$1332 (black histogram). The continuous red line is
the best-fitting Voigt profile model for an \HI\ column density
$\log N({\rm H\,\textsc{i}})/{\rm cm}^{-2}=20.39\pm0.04$.
The remaining panels display a selection of the associated
metal absorption lines, with the best-fitting model overplotted.
Fitted line blends are shown by the blue lines.
The red tick marks above the spectra indicate the position of
the absorption component. Note the different y-axis scale used
to expose the weak \CrII, \NiII, and \ZnII\ lines, as well as
\FeII\,$\lambda2260$. For \ZnII\,$\lambda2026$, we overplot
the cloud model for the derived $3\sigma$ upper limit. The
normalized quasar continuum and zero-level are shown by
the horizontal blue long-dashed and green dotted lines respectively.
The red wing of the \Lya\ profile is slightly depressed relative to
the QSO continuum due to another DLA along this sightline at
$z_{\rm abs}\simeq2.38223$.
  }
  \label{fig:J1111p1332_al}
\end{figure*}

\citet{Coo13} included the Fe-peak element ratios in this DLA in their
study.
Here, we report a detailed profile analysis which includes all of
the available metal species. In total, we detect the absorption lines for
ten elements and place an interesting upper limit on the Zn abundance.
The absorption profiles are very well fit by a single absorption component
centered at redshift $z_{\rm abs}=2.270940\pm0.000001$
with turbulent Doppler parameter $b_{\rm turb}=2.2\pm0.1\,{\rm km\,\,s}^{-1}$ and gas kinetic
temperature $T_{\rm gas}=13000\pm2000$ K. 
We reproduce the data and our best-fitting model for
this system in Figure~\ref{fig:J1111p1332_al} where the top panel 
shows the damped \Lya\ line profile, and
best-fitting model for $\log N({\rm H\,\textsc{i}})/{\rm cm}^{-2}=20.39\pm0.04$.
The remaining panels are a selection of  metal absorption lines. 
The y-axis scale has been reduced in the six panels showing the weakest
absorption lines. Column densities for all
detected elements and ion stages in this system are collected in
Table~\ref{tab:J1111p1332_cd}.

\begin{thebibliography}{99}

\bibitem[\protect\citeauthoryear{Akerman et al.}{2005}]{Ake05}
Akerman C.~J., Ellison S.~L., Pettini M., Steidel C.~C., 2005, A\&A, 440, 499

\bibitem[\protect\citeauthoryear{Alavi et al.}{2014}]{Ala14}
Alavi A., et al., 2014, ApJ, 780, 143

\bibitem[\protect\citeauthoryear{Amor{\'{\i}}n et al.}{2014}]{Amo14}
Amor{\'{\i}}n R., et al., 2014, ApJL, 788, L4 

\bibitem[\protect\citeauthoryear{Aoki et al.}{2007}]{Aok07}
Aoki W., Beers T.~C., Christlieb N., Norris J.~E., Ryan S.~G., Tsangarides S., 2007, ApJ, 655, 492

\bibitem[\protect\citeauthoryear{Asplund et al.}{2009}]{Asp09}
Asplund M., Grevesse N., Sauval A.~J., Scott P., 2009, ARA\&A, 47, 481

\bibitem[\protect\citeauthoryear{Barnes \& Haehnelt}{2014}]{BarHae14}
Barnes L.~A., Haehnelt M.~G., 2014, MNRAS, 440, 2313

\bibitem[\protect\citeauthoryear{Belokurov et al.}{2006}]{Bel06} 
Belokurov, V.,  Zucker, D.~B., Evans, N.~W., et al.\ 2006, ApJL, 647, L111 

\bibitem[\protect\citeauthoryear{Belokurov}{2013}]{Bel13}
Belokurov V., 2013, NewAR, 57, 100

\bibitem[\protect\citeauthoryear{Bensby et al.}{2005}]{Ben05}
Bensby T., Feltzing S., Lundstr{\"o}m I., Ilyin I., 2005, A\&A, 433, 185

\bibitem[\protect\citeauthoryear{Bernstein-Cooper et al.}{2014}]{Ber14}
Bernstein-Cooper E.~Z., et al., 2014, arXiv, arXiv:1404.5298

\bibitem[\protect\citeauthoryear{Bird et al.}{2013}]{Bir13}
Bird S., Vogelsberger M., Sijacki D., Zaldarriaga M., Springel V., Hernquist L., 2013, MNRAS, 429, 3341

\bibitem[\protect\citeauthoryear{Bird et al.}{2015}]{Bir15}
Bird S., Haehnelt M., Neeleman M., Genel S., Vogelsberger M., Hernquist L., 2015, MNRAS, 447, 1834

\bibitem[\protect\citeauthoryear{Bonifacio et al.}{2000}]{Bon00}
Bonifacio P., Hill V., Molaro P., Pasquini L., Di Marcantonio P., Santin P., 2000, A\&A, 359, 663

\bibitem[\protect\citeauthoryear{Bonifacio et al.}{2004}]{Bon04}
Bonifacio P., Sbordone L., Marconi G., Pasquini L., Hill V., 2004, A\&A, 414, 503

\bibitem[\protect\citeauthoryear{Bromm \& Yoshida}{2011}]{BroYos11}
Bromm V., Yoshida N., 2011, ARA\&A, 49, 373

\bibitem[\protect\citeauthoryear{Bullock, Kravtsov, \& Weinberg}{2000}]{BulKraWei00}
Bullock J.~S., Kravtsov A.~V., Weinberg D.~H., 2000, ApJ, 539, 517

\bibitem[\protect\citeauthoryear{Bunker et al.}{1999}]{Bun99}
Bunker A.~J., Warren S.~J., Clements D.~L., Williger G.~M., Hewett P.~C., 1999, MNRAS, 309, 875

\bibitem[\protect\citeauthoryear{Cantalupo, Lilly, \& Haehnelt}{2012}]{CanLilHae12}
Cantalupo S., Lilly S.~J., Haehnelt M.~G., 2012, MNRAS, 425, 1992

\bibitem[\protect\citeauthoryear{Carswell et al.}{2012}]{Car12}
Carswell R.~F., Becker G.~D., Jorgenson R.~A., Murphy M.~T., Wolfe A.~M., 2012, MNRAS, 422, 1700

\bibitem[\protect\citeauthoryear{Cen}{2012}]{Cen12}
Cen R., 2012, ApJ, 748, 121

\bibitem[\protect\citeauthoryear{Chieffi \& Limongi}{2004}]{ChiLim04}
Chieffi A., Limongi M., 2004, ApJ, 608, 405

\bibitem[\protect\citeauthoryear{Christensen et al.}{2009}]{Chr09}
Christensen L., Noterdaeme P., Petitjean P., Ledoux C., Fynbo J.~P.~U., 2009, A\&A, 505, 1007

\bibitem[\protect\citeauthoryear{Christensen et al.}{2014}]{Chr14}
Christensen L., M{\o}ller P., Fynbo J.~P.~U., Zafar T., 2014, MNRAS, 445, 225

\bibitem[\protect\citeauthoryear{Cohen \& Huang}{2009}]{CohHua09}
Cohen J.~G., Huang W., 2009, ApJ, 701, 1053

\bibitem[\protect\citeauthoryear{Cohen \& Huang}{2010}]{CohHua10}
Cohen J.~G., Huang W., 2010, ApJ, 719, 931

\bibitem[\protect\citeauthoryear{Cooke et al.}{2006}]{Coo06}
Cooke J., Wolfe A.~M., Gawiser E., Prochaska J.~X., 2006, ApJ, 652, 994

\bibitem[\protect\citeauthoryear{Cooke et al.}{2011a}]{Coo11a}
Cooke R., Pettini M., Steidel C.~C., Rudie G.~C., Jorgenson R.~A., 2011a, MNRAS, 412, 1047

\bibitem[\protect\citeauthoryear{Cooke et al.}{2011b}]{Coo11b}
Cooke R., Pettini M., Steidel C.~C., Rudie G.~C., Nissen P.~E., 2011b, MNRAS, 417, 1534

\bibitem[\protect\citeauthoryear{Cooke, Pettini,  \& Murphy}{2012}]{Coo12}
Cooke R., Pettini M., Murphy M.~T., 2012, MNRAS, 425, 347

\bibitem[\protect\citeauthoryear{Cooke et al.}{2013}]{Coo13}
Cooke R., Pettini M., Jorgenson R.~A., Murphy M.~T., Rudie G.~C., Steidel C.~C., 2013, MNRAS, 431, 1625

\bibitem[\protect\citeauthoryear{Cooke et al.}{2014}]{Coo14}
Cooke R.~J., Pettini M., Jorgenson R.~A., Murphy M.~T., Steidel C.~C., 2014, ApJ, 781, 31

\bibitem[\protect\citeauthoryear{Cooke \& Madau}{2014}]{CooMad14}
Cooke R., Madau P., 2014, ApJ submitted, arXiv:1405.7369

\bibitem[\protect\citeauthoryear{de Boer et al.}{2014}]{deB14}
de Boer T.~J.~L., Belokurov V., Beers T.~C., Lee Y.~S., 2014, arXiv:1406.3352

\bibitem[\protect\citeauthoryear{Dutta et al.}{2014}]{Dut14}
Dutta R., Srianand R., Rahmani H., Petitjean P., Noterdaeme P., Ledoux C., 2014, arXiv, arXiv:1402.2975

\bibitem[\protect\citeauthoryear{Ellison et al.}{2010}]{Ell10}
Ellison S.~L., Prochaska J.~X., Hennawi J., Lopez S., Usher C., Wolfe A.~M., Russell D.~M., Benn C.~R., 2010, MNRAS, 406, 1435

\bibitem[\protect\citeauthoryear{Ferland et al.}{2013}]{Fer13}
Ferland G.~J., et al., 2013, Rev. Mex. Astron. Astrofis., 49, 137

\bibitem[\protect\citeauthoryear{Font-Ribera et al.}{2012}]{Fon12}
Font-Ribera A., et al., 2012, JCAP, 11, 59

\bibitem[\protect\citeauthoryear{Frebel \& Bromm}{2012}]{FreBro12}
Frebel A., Bromm V., 2012, ApJ, 759, 115

\bibitem[\protect\citeauthoryear{Frebel, Simon, \& Kirby}{2014}]{FreSimKir14}
Frebel A., Simon J.~D., Kirby E.~N., 2014, arXiv, arXiv:1403.6116

\bibitem[\protect\citeauthoryear{Fumagalli et al.}{2011}]{Fum11}
Fumagalli M., Prochaska J.~X., Kasen D., Dekel A., Ceverino D., Primack J.~R., 2011, MNRAS, 418, 1796

\bibitem[\protect\citeauthoryear{Fumagalli et al.}{2014}]{Fum14}
Fumagalli M., O'Meara J.~M., Prochaska J.~X., Kanekar N., Wolfe A.~M., 2014, MNRAS, 444, 1282

\bibitem[\protect\citeauthoryear{Fynbo et al.}{2010}]{Fyn10} 
Fynbo J.~P.~U., et al., 2010, MNRAS, 408, 2128 

\bibitem[\protect\citeauthoryear{Gilmore et al.}{2013}]{Gil13}
Gilmore G., Norris J.~E., Monaco L., Yong D., Wyse R.~F.~G., Geisler D., 2013, ApJ, 763, 61

\bibitem[\protect\citeauthoryear{Giovanelli et al.}{2013}]{Gio13}
Giovanelli R., et al., 2013, AJ, 146, 15

\bibitem[\protect\citeauthoryear{Haardt \& Madau}{2012}]{HarMad12}
Haardt F., Madau P., 2012, ApJ, 746, 125

\bibitem[\protect\citeauthoryear{Haehnelt, Steinmetz, \& Rauch}{1998}]{HaeSteRau98}
Haehnelt M.~G., Steinmetz M., Rauch M., 1998, ApJ, 495, 647

\bibitem[\protect\citeauthoryear{Heger \& Woosley}{2010}]{HegWoo10}
Heger A., Woosley S.~E., 2010, ApJ, 724, 341

\bibitem[\protect\citeauthoryear{Heiles \& Troland}{2003}]{HeiTro03}
Heiles C., Troland T.~H., 2003, ApJ, 586, 1067

\bibitem[\protect\citeauthoryear{Hendricks et al.}{2014}]{Hen14}
Hendricks B., Koch A., Lanfranchi G.~A., Boeche C., Walker M., Johnson C.~I., Penarrubia J., Gilmore G., 2014, arXiv, arXiv:1402.6328

\bibitem[\protect\citeauthoryear{Hill, Barbuy, \& Spite}{1997}]{HilBarSpi97}
Hill V., Barbuy B., Spite M., 1997, A\&A, 323, 461

\bibitem[\protect\citeauthoryear{Hill}{2012}]{Hil12}
Hill, V., \&\ DART Collaboration 2012, in ASP Conf. Ser. 458, Galactic Archaeology: Near-Field Cosmology and the Formation of the Milky Way, ed. W. Aoki, M. Ishigaki, T. Suda, T. Tsujimoto, and N. Arimoto (San Francisco, CA: ASP), 297

\bibitem[\protect\citeauthoryear{Howk, Wolfe, \& Prochaska}{2005}]{How05}
Howk J.~C., Wolfe A.~M., Prochaska J.~X., 2005, ApJ, 622, L81

\bibitem[\protect\citeauthoryear{Illingworth et al.}{2013}]{Ill13}
Illingworth, G.~D., Magee, D., Oesch, P.~A., et al.\ 2013, ApJS, 209, 6

\bibitem[\protect\citeauthoryear{Jenkins \& Tripp}{2001}]{JenTri01}
Jenkins, E.~B., \& Tripp, T.~M.\ 2001, ApJSupp,  137, 297 

\bibitem[\protect\citeauthoryear{Jenkins \& Tripp}{2006}]{JenTri06}
Jenkins, E.~B., \& Tripp, T.~M. 2006, ApJ, 637, 548

\bibitem[\protect\citeauthoryear{Jenkins \& Tripp}{2011}]{JenTri11}
Jenkins, E.~B., \& Tripp, T.~M.\ 2011, ApJ,  734, 65

\bibitem[\protect\citeauthoryear{Jorgenson et al.}{2009}]{Jor09}
Jorgenson R.~A., Wolfe A.~M., Prochaska J.~X., Carswell R.~F., 2009, ApJ, 704, 247

\bibitem[\protect\citeauthoryear{Jorgenson, Wolfe, \& Prochaska}{2010}]{Jor10}
Jorgenson R.~A., Wolfe A.~M., Prochaska J.~X., 2010, ApJ, 722, 460

\bibitem[\protect\citeauthoryear{Jorgenson, Murphy, \& Thompson}{2013}]{JorMurTho13}
Jorgenson R.~A., Murphy M.~T., Thompson R., 2013, MNRAS, 435, 482

\bibitem[\protect\citeauthoryear{Jorgenson \& Wolfe}{2014}]{JorWol14}
Jorgenson R.~A., Wolfe A.~M., 2014, ApJ, 785, 16

\bibitem[\protect\citeauthoryear{Jorgenson et al.}{2014}]{Jor14}
Jorgenson R.~A., Murphy M.~T., Thompson R., Carswell R.~F., 2014, MNRAS, 443, 2783

\bibitem[\protect\citeauthoryear{Kanekar \& Chengalur}{2001}]{KanChe01}
Kanekar N., Chengalur J.~N., 2001, A\&A, 369, 42

\bibitem[\protect\citeauthoryear{Kanekar et al.}{2014}]{Kan14}
Kanekar N., et al., 2014, MNRAS, 438, 2131

\bibitem[\protect\citeauthoryear{Kirby et al.}{2011a}]{Kir11a}
Kirby E.~N., Lanfranchi G.~A., Simon J.~D., Cohen J.~G., Guhathakurta P., 2011a, ApJ, 727, 78

\bibitem[\protect\citeauthoryear{Kirby et al.}{2011b}]{Kir11b}
Kirby E.~N., Cohen J.~G., Smith G.~H., Majewski S.~R., Sohn S.~T., Guhathakurta P., 2011b, ApJ, 727, 79

\bibitem[\protect\citeauthoryear{Koch et al.}{2008}]{Koc08}
Koch A., Grebel E.~K., Gilmore G.~F., Wyse R.~F.~G., Kleyna J.~T., Harbeck D.~R., Wilkinson M.~I., Wyn Evans N., 2008, AJ, 135, 1580

\bibitem[\protect\citeauthoryear{Kowal, Lazarian, \& Beresnyak}{2007}]{KowLazBer07}
Kowal G., Lazarian A., Beresnyak A., 2007, ApJ, 658, 423

\bibitem[\protect\citeauthoryear{Krogager et al.}{2012}]{Kro12}
Krogager J.-K., Fynbo J.~P.~U., M{\o}ller P., Ledoux C., Noterdaeme P., Christensen L., Milvang-Jensen B., Sparre M., 2012, MNRAS, 424, L1

\bibitem[\protect\citeauthoryear{Krogager et al.}{2013}]{Kro13}
Krogager, J.-K., Fynbo, J. P. U., Ledoux, C., et al. 2013, MNRAS, 433, 3091

\bibitem[\protect\citeauthoryear{Kulkarni et al.}{2000}]{Kul00}
Kulkarni V.~P., Hill J.~M., Schneider G., Weymann R.~J., Storrie-Lombardi L.~J., Rieke M.~J., Thompson R.~I., Jannuzi B.~T., 2000, ApJ, 536, 36

\bibitem[\protect\citeauthoryear{Lai et al.}{2011}]{Lai11}
Lai D.~K., Lee Y.~S., Bolte M., Lucatello S., Beers T.~C., Johnson J.~A., Sivarani T., Rockosi C.~M., 2011, ApJ, 738, 51

\bibitem[\protect\citeauthoryear{Ledoux et al.}{2006}]{Led06}
Ledoux, C., Petitjean, P., Fynbo, J.~P.~U., M{\o}ller, P., \& Srianand, R.\ 2006, A\&A, 457, 71

\bibitem[\protect\citeauthoryear{Letarte et al.}{2010}]{Let10}
Letarte B., et al., 2010, A\&A, 523, A17

\bibitem[\protect\citeauthoryear{Liszt}{2001}]{Lis01}
Liszt H., 2001, A\&A, 371, 698

\bibitem[\protect\citeauthoryear{Lowenthal et al.}{1995}]{Low95}
Lowenthal J.~D., Hogan C.~J., Green R.~F., Woodgate B., Caulet A., Brown L., Bechtold J., 1995, ApJ, 451, 484

\bibitem[\protect\citeauthoryear{Mannucci, Della Valle, \& Panagia}{2006}]{ManDelPan06}
Mannucci F., Della Valle M., Panagia N., 2006, MNRAS, 370, 773

\bibitem[\protect\citeauthoryear{Maio, Tescari, \& Cooke}{2014}]{MaiTesCoo14}
Maio U., Tescari E., Cooke R., 2014, MNRAS submitted

\bibitem[\protect\citeauthoryear{Markwardt}{2009}]{Mar09}
Markwardt C.~B., 2009, ASPC, 411, 251

\bibitem[\protect\citeauthoryear{Mateo}{1998}]{Mat98}
Mateo M.~L., 1998, ARA\&A, 36, 435

\bibitem[\protect\citeauthoryear{McConnachie}{2012}]{McC12}
McConnachie A.~W., 2012, AJ, 144, 4

\bibitem[\protect\citeauthoryear{McKee \& Ostriker}{2007}]{McKOst07}
McKee C.~F., Ostriker E.~C., 2007, ARA\&A, 45, 565

\bibitem[\protect\citeauthoryear{M{\o}ller et al.}{2013}]{Mol13}
M{\o}ller P., Fynbo J.~P.~U., Ledoux C., Nilsson K.~K., 2013, MNRAS, 430, 2680

\bibitem[\protect\citeauthoryear{Monaco et al.}{2005}]{Mon05}
Monaco L., Bellazzini M., Bonifacio P., Ferraro F.~R., Marconi G., Pancino E., Sbordone L., Zaggia S., 2005, A\&A, 441, 141

\bibitem[\protect\citeauthoryear{Morton}{2003}]{Mor03}
Morton, D.~C. 2003, ApJS, 149, 205

\bibitem[\protect\citeauthoryear{Morton et al.}{1973}]{Mor73}
Morton, D.~C., Drake, J.~F., Jenkins, E.~B., et al.\ 1973, ApJL, 181, L103

\bibitem[\protect\citeauthoryear{Murphy et al.}{2007}]{Mur07}
Murphy, M.~T., Curran, S.~J., Webb, J.~K., M{\'e}nager, H., \& Zych, B.~J.\ 2007, MNRAS, 376, 673

\bibitem[\protect\citeauthoryear{Murphy \& Berengut}{2014}]{MurBer14}
Murphy M.~T., Berengut J.~C., 2014, MNRAS, 438, 388

\bibitem[\protect\citeauthoryear{Neeleman et al.}{2013}]{Nee13}
Neeleman M., Wolfe A.~M., Prochaska J.~X., Rafelski M., 2013, ApJ, 769, 54

\bibitem[\protect\citeauthoryear{Norris et al.}{2010}]{Nor10}
Norris J.~E., Wyse R.~F.~G., Gilmore G., Yong D., Frebel A., Wilkinson M.~I., Belokurov V., Zucker D.~B., 2010, ApJ, 723, 1632

\bibitem[\protect\citeauthoryear{Noterdaeme et al.}{2012}]{Not12}
Noterdaeme P., et al., 2012, A\&A, 540, A63

\bibitem[\protect\citeauthoryear{Penprase et al.}{2010}]{Pen10}
Penprase B.~E., Prochaska J.~X., Sargent W.~L.~W., Toro-Martinez I., Beeler D.~J., 2010, ApJ, 721, 1

\bibitem[\protect\citeauthoryear{P{\'e}roux et al.}{2012}]{Per12}
P{\'e}roux, C.,  Bouch{\'e}, N., Kulkarni, V.~P., York, D.~G., 
\& Vladilo, G.\ 2012, MNRAS, 419, 3060 

\bibitem[\protect\citeauthoryear{Petitjean, Srianand, \& Ledoux}{2000}]{PetSriLed00}
 Petitjean, P., Srianand, R., \& Ledoux, C.\ 2000, A\&A, 364, L26 

\bibitem[\protect\citeauthoryear{Pettini et al.}{1983}]{Pet83} 
Pettini, M., Hunstead, 
R.~W., Murdoch, H.~S., \& Blades, J.~C.\ 1983, ApJ, 273, 436

\bibitem[\protect\citeauthoryear{Pettini et al.}{1997}]{Pet97}
Pettini M., King D.~L., Smith L.~J., Hunstead R.~W., 1997, ApJ, 478, 536

\bibitem[\protect\citeauthoryear{Pettini \& Cooke}{2012}]{PetCoo12}
Pettini M. \& Cooke R., 2012, MNRAS, 425, 2477

\bibitem[\protect\citeauthoryear{Pettini et al.}{2008}]{Pet08}
Pettini M., Zych B.~J., Steidel C.~C., Chaffee F.~H., 2008, MNRAS, 385, 2011

\bibitem[\protect\citeauthoryear{Pomp{\'e}ia et al.}{2008}]{Pom08}
Pomp{\'e}ia L., et al., 2008, A\&A, 480, 379

\bibitem[\protect\citeauthoryear{Pontzen et al.}{2008}]{Pon08}
Pontzen A., et al., 2008, MNRAS, 390, 1349

\bibitem[\protect\citeauthoryear{Prochaska et al.}{2008}]{Pro08}
Prochaska, J.~X., Chen, H.-W., Wolfe, A.~M., Dessauges-Zavadsky, M., \& Bloom, J.~S.\ 2008, ApJ, 672, 59

\bibitem[\protect\citeauthoryear{Prochaska et al.}{2003}]{Pro03}
Prochaska J.~X., Gawiser E., Wolfe A.~M., Cooke J., Gelino D., 2003, ApJS, 147, 227

\bibitem[\protect\citeauthoryear{Prochaska \& Wolfe}{1997}]{ProWol97}
Prochaska J.~X., Wolfe A.~M., 1997, ApJ, 487, 73

\bibitem[\protect\citeauthoryear{Prochaska \& Wolfe}{2010}]{ProWol10}
Prochaska J.~X., Wolfe A.~M., 2010, arXiv, arXiv:1009.3960

\bibitem[\protect\citeauthoryear{Rafelski et al.}{2012}]{Raf12}
Rafelski M., Wolfe A.~M., Prochaska J.~X., Neeleman M., Mendez A.~J., 2012, ApJ, 755, 89

\bibitem[\protect\citeauthoryear{Rahmati \& Schaye}{2014}]{RahSch14}
Rahmati A., Schaye J., 2014, MNRAS, 438, 529

\bibitem[\protect\citeauthoryear{Rauch et al.}{2008}]{Rau08}
Rauch M., et al., 2008, ApJ, 681, 856

\bibitem[\protect\citeauthoryear{Redfield \& Linsky}{2004}]{RedLin04}
Redfield S., Linsky J.~L., 2004, ApJ, 613, 1004

\bibitem[\protect\citeauthoryear{Ryan et al.}{2005}]{Rya05}
Ryan S.~G., Aoki W., Norris J.~E., Beers T.~C., 2005, ApJ, 635, 349

\bibitem[\protect\citeauthoryear{Ryan-Weber et al.}{2008}]{Rya08}
Ryan-Weber E.~V., Begum A., Oosterloo T., Pal S., Irwin M.~J., Belokurov V., Evans N.~W., Zucker D.~B., 2008, MNRAS, 384, 535

\bibitem[\protect\citeauthoryear{Saito et al.}{2009}]{Sai09}
Saito Y.-J., Takada-Hidai M., Honda S., Takeda Y., 2009, PASJ, 61, 549

\bibitem[\protect\citeauthoryear{Salvadori \& Ferrara}{2012}]{SalFer12}
Salvadori S., Ferrara A., 2012, MNRAS, 421, L29

\bibitem[\protect\citeauthoryear{Sbordone et al.}{2007}]{Sbo07}
Sbordone L., Bonifacio P., Buonanno R., Marconi G., Monaco L., Zaggia S., 2007, A\&A, 465, 815

\bibitem[\protect\citeauthoryear{Shetrone, Bolte, \& Stetson}{1998}]{SheBolSte98}
Shetrone M.~D., Bolte M., Stetson P.~B., 1998, AJ, 115, 1888

\bibitem[\protect\citeauthoryear{Shetrone, C{\^o}t{\'e}, \& Sargent}{2001}]{SheCotSar01}
Shetrone M.~D., C{\^o}t{\'e} P., Sargent W.~L.~W., 2001, ApJ, 548, 592

\bibitem[\protect\citeauthoryear{Skillman et al.}{2013}]{Ski13}
Skillman E.~D., et al., 2013, AJ, 146, 3

\bibitem[\protect\citeauthoryear{Srianand et al.}{2005}]{Sri05}
Srianand, R., Petitjean, P., Ledoux, C., Ferland, G., \& Shaw, G.\ 2005, MNRAS, 362, 549

\bibitem[\protect\citeauthoryear{Srianand et al.}{2012}]{Sri12}
Srianand R., Gupta N., Petitjean P., Noterdaeme P., Ledoux C., Salter C.~J., Saikia D.~J., 2012, MNRAS, 421, 651

\bibitem[\protect\citeauthoryear{Starkenburg et al.}{2013}]{Sta13}
Starkenburg E., et al., 2013, A\&A, 549, A88

\bibitem[\protect\citeauthoryear{Tescari et al.}{2009}]{Tes09}
Tescari E., Viel M., Tornatore L., Borgani S., 2009, MNRAS, 397, 411

\bibitem[\protect\citeauthoryear{Tinsley}{1979}]{Tin79}
Tinsley B.~M., 1979, ApJ, 229, 1046

\bibitem[\protect\citeauthoryear{Tolstoy, Hill, \& Tosi}{2009}]{TolHilTos09}
Tolstoy E., Hill V., Tosi M., 2009, ARA\&A, 47, 371

\bibitem[\protect\citeauthoryear{Tolstoy \& Saha}{1996}]{TolSah96}
Tolstoy E., Saha A., 1996, ApJ, 462, 672

\bibitem[\protect\citeauthoryear{Tumlinson et al.}{2010}]{Tum10}
Tumlinson J., et al., 2010, ApJ, 718, L156

\bibitem[\protect\citeauthoryear{Vallerga}{1996}]{Val96}
Vallerga J., 1996, SSRv, 78, 277

\bibitem[\protect\citeauthoryear{van de Voort et al.}{2012}]{van12}
van de Voort F., Schaye J., Altay G., Theuns T., 2012, MNRAS, 421, 2809

\bibitem[\protect\citeauthoryear{Vargas, Geha, \& Tollerud}{2014}]{Var14}
Vargas L.~C., Geha M.~C., Tollerud E.~J., 2014, arXiv, arXiv:1406.0510

\bibitem[\protect\citeauthoryear{Venn}{1999}]{Ven99}
Venn K.~A., 1999, ApJ, 518, 405

\bibitem[\protect\citeauthoryear{Venn et al.}{2004}]{Ven04}
Venn K.~A., Irwin M., Shetrone M.~D., Tout C.~A., Hill V., Tolstoy E., 2004, AJ, 128, 1177

\bibitem[\protect\citeauthoryear{Vladilo et al.}{2011}]{Vla11}
Vladilo G., Abate C., Yin J., Cescutti G., Matteucci F., 2011, A\&A, 530, A33

\bibitem[\protect\citeauthoryear{Vogt et al.}{1994}]{Vog94}
Vogt S.~S., et al., 1994, SPIE, 2198, 362

\bibitem[\protect\citeauthoryear{Walker et al.}{2013}]{Wal13}
Walker~M., 2013, Planets, Stars and Stellar Systems. Vol. 5: Galactic Structure and Stellar Populations (Dordrecht: Springer), 1039

\bibitem[\protect\citeauthoryear{Weisz et al.}{2011}]{Wei11}
Weisz D.~R., et al., 2011, ApJ, 739, 5

\bibitem[\protect\citeauthoryear{Weisz et al.}{2014a}]{Wei14a}
Weisz D.~R., et al., 2014a, ApJ submitted, arXiv:1404.7144

\bibitem[\protect\citeauthoryear{Weisz et al.}{2014b}]{Wei14b}
Weisz D.~R., Dolphin A.~E., Skillman E.~D., Holtzman J., Gilbert K.~M., Dalcanton J.~J., Williams B.~F., 2014b, ApJ submitted, arXiv:1405.3281

\bibitem[\protect\citeauthoryear{Wheeler, Sneden, \& Truran}{1989}]{WheSneTru89}
Wheeler J.~C., Sneden C., Truran J.~W., Jr., 1989, ARA\&A, 27, 279

\bibitem[\protect\citeauthoryear{Willman et al.}{2005a}]{Wil05a}
Willman B., et al., 2005a, AJ, 129, 2692

\bibitem[\protect\citeauthoryear{Willman et al.}{2005b}]{Wil05b}
Willman B., et al., 2005b, ApJ, 626, L85

\bibitem[\protect\citeauthoryear{Wolfe, Gawiser, \& Prochaska}{2005}]{WolGawPro05}
Wolfe A.~M., Gawiser E., Prochaska J.~X., 2005, ARA\&A, 43, 861

\bibitem[\protect\citeauthoryear{Wolfe, Prochaska, \& Gawiser}{2003}]{WolProGaw03} 
Wolfe, A.~M., Prochaska, J.~X., \& Gawiser, E.\ 2003, ApJ, 593, 215 

\bibitem[\protect\citeauthoryear{Wolfe et al.}{1986}]{Wol86}
Wolfe A.~M., Turnshek D.~A., Smith H.~E., Cohen R.~D., 1986, ApJS, 61, 249

\bibitem[\protect\citeauthoryear{Wolfire et al.}{1995}]{Wol95}
Wolfire M.~G., Hollenbach D., McKee C.~F., Tielens A.~G.~G.~M., Bakes E.~L.~O., 1995, ApJ, 443, 152

\bibitem[\protect\citeauthoryear{Wolfire et al.}{2003}]{Wol03}
Wolfire, M.~G., McKee, 
C.~F., Hollenbach, D., \& Tielens, A.~G.~G.~M.\ 2003, ApJ,  587, 278 

\bibitem[\protect\citeauthoryear{Woosley \& Weaver}{1995}]{WooWea95}
Woosley S.~E., Weaver T.~A., 1995, ApJS, 101, 181

\bibitem[\protect\citeauthoryear{York et al.}{2000}]{Yor00}
York D.~G., et al., 2000, AJ, 120, 1579

\end{thebibliography}
\end{document}